\documentclass[12pt]{article}\usepackage[hyperfootnotes=false]{hyperref}
\usepackage{epsfig}
\usepackage{float}
\usepackage{empheq}
\usepackage{bbold}

\usepackage[utf8]{inputenc}
\usepackage{amsmath}

\usepackage{caption}

\usepackage{amsmath}
\usepackage{amssymb}
\usepackage{graphicx}
\newlength{\abstractwidth}
\renewcommand{\thefootnote}{\fnsymbol{footnote}}
\renewcommand{\thanks}[1]{\footnote{#1}} 
\newcommand{\starttext}{
\setcounter{footnote}{0}
\renewcommand{\thefootnote}{\arabic{footnote}}}

\newcommand{\be}{\begin{equation}}
\newcommand{\bea}{\begin{eqnarray}}
\newcommand{\eea}{\end{eqnarray}}
\newcommand{\beq}{\begin{equation}}
\newcommand{\ee}{\end{equation}}

\newcommand*\widefbox[1]{\fbox{\hspace{2em}#1\hspace{2em}}}

\def\dss{de Sitter space}
\def\dsp.{de Sitter space.}
\def\eq{&=&}

\def\la{\langle}
\def\ra{\rangle}
\def\simleq{\; \raise0.3ex\hbox{$<$\kern-0.75em
\raise-1.1ex\hbox{$\sim$}}\; }
\def\simgeq{\; \raise0.3ex\hbox{$>$\kern-0.75em
\raise-1.1ex\hbox{$\sim$}}\; }

\def\bi{\begin{itemize}}
\def\ei{\end{itemize}}
\def\S{Schwarzschild}
\def\sc{\setcounter{equation}{0}}
\def\dof{degrees of freedom }
\def\CA{{\cal{A}}}

\def\CH{{\cal{H}}}

\def\CJ{{\cal{J}}}

\def\CT{{\cal{T}}}

\def\t{\tau}

\def\bsub{ \begin{subequations}
\begin{empheq}[box=\widefbox]{align}  }
\def\esub{ \end{empheq}
\end{subequations}}

\def\1{\(  \mathbb{1} \)}

   \def\l{L_{c}}

 \def\lf{\left(}
    \def\rg{\right)}

  \def\bn{\bigskip \noindent}

 \def\bm{\begin{bmatrix}}
 \def\em{\end{bmatrix}}

    \def\dk{${\rm DSSYK_{\infty}}$}
    \def\qnm{quasinormal modes}

\makeatletter
\g@addto@macro\normalsize{%
  \setlength\abovedisplayskip{10pt}
  \setlength\belowdisplayskip{20pt}
  \setlength\abovedisplayshortskip{10pt}
  \setlength\belowdisplayshortskip{20pt}
}
\makeatother

\usepackage{color}

\begin{document}

\begin{titlepage}


 \rightline{}
\bigskip
\bigskip\bigskip\bigskip\bigskip
\bigskip

\centerline{\Large \bf {De Sitter Space, Double-Scaled SYK,  and }}

\bn

\centerline{\Large \bf {the Separation of Scales in the Semiclassical Limit }}

\bigskip
\begin{center}
	\bf   Leonard Susskind \rm

\bigskip

 Stanford Institute for Theoretical Physics and Department of Physics, \\
Stanford University,
Stanford, CA 94305-4060, USA \\

\end{center}

\bn


\begin{abstract}

In the semiclassical limit of de Sitter gravity a separation of scales takes place that divides the theory into a ``cosmic" sector and a ``microscopic" sector. A similar separation takes place in the double-scaled limit of SYK theory.  We examine the scaling behaviors that accompany these limits and find parallels that support the  previously conjectured duality between Jackiw-Teitelboim gravity (with positive cosmological constant), and double-scaled SYK.

This paper is a companion to  "dS JT Gravity and Double-Scaled SYK" by Adel Rahman, to appear simultaneously with this paper.

\end{abstract}

\end{titlepage}



 \rightline{}
\bigskip
\bigskip\bigskip\bigskip\bigskip
\bigskip





\bn


\bn

\starttext \baselineskip=17.63pt \setcounter{footnote}{0}

\tableofcontents

\Large


\sc
\section{Introduction}

Imagine a planet whose surface-gravity,  people, houses, cities and other local surface features are  similar to those of  the Earth’s, but the diameter of the planet is a tunable parameter that can be taken arbitrarily large. In the limit of infinite diameter is the surface flat? Or is it spherical? 

That depends on the questions one asks.
If we are interested in people, houses and cities then for practical purposes the geometry approaches the ``flat-space limit.” However, if we are interested in the properties of triangles whose sides are a fixed fraction of the planet's diameter then the geometry is spherical. But in the limit of infinite diameter the separation of scales becomes infinite and the two aspects of the geometry  decouple.

 The situation calls for two distinct systems of units; one in which lengths are measured in ``microscopic” units such as feet, meters, or miles, with the sizes of people and cities being fixed and the diameter of the planet becoming infinite; and one in which the diameter is fixed but people and cities become infinitesimal. In the limit we need both in order to describe the local and the global geometry, and to give meaning to whether the geometry is spherical or  flat.
 
      Exactly the same thing happens in the semiclassical limit\footnote{``Semiclassical" has been used to mean different things by different authors. In this paper we follow  \cite{Witten:2021jzq}\cite{Chandrasekaran:2022cip}:  by the SCL we mean the limit in which the entropy of the de Sitter space goes to infinity.} (SCL) of de Sitter space. The SCL is one in which the entropy, and therefore the area of the horizon, diverges but ``micro” length scales stay fixed\footnote{The term micro will mean anything small enough to be insensitive to the global de Sitter radius.}. Such micro-scales include the Planck length, the radius of an atom, or the size of a galaxy. For practical purposes microscopic physics might as well be taking place in flat space.
      But other properties of the universe will be sensitive to the cosmic geometry, for example the behavior of quasinormal modes and the existence of very low frequency Gibbons-Hawking radiation \cite{Gibbons:1977mu}. Understanding the full local and global geometry requires two sets of units---microscopic units and cosmic units.

 According to a recent conjecture \cite{Susskind:2021esx}\cite{Susskind:2022dfz}\cite{Lin:2022nss}\cite{Adel} 
  there is a holographic duality between a static patch of Jackiw-Teitelboim de Sitter gravity, and the double-scaled limit of  SYK
  \cite{Cotler:2016fpe}\cite{Berkooz:2018jqr}\cite{Lin:2022rbf} 
  at infinite temperature (\dk).
 This paper is about the separation of scales which takes place
  in the SCL of de Sitter space, and how it relates  to a similar separation of scales in the $N\to \infty$ limit of \dk.

\bn

By  microscopic (or just micro)  I will mean not only Planckian but all phenomena which track the Planck scale: elementary particles; atomic physics; chemistry; and even sub-cosmological astronomy. These are identified by their insensitivity to the cosmological scale, and they have a flat-space limit.

  By contrast there are those things which are sensitive to the cosmological scale but not to microscopic details. These include (in cosmological units) the decay time for \qnm \ (QNM), the energy of a Hawking quantum in a static patch, some late-time features of scrambling, and more. In the SCL we can choose units  to follow either the  micro or the cosmic phenomena but not both. If we choose to measure time in Planck units we can follow elementary particle phenomena, but QNMs will have infinite lifetimes; if we measure time in cosmic units we can follow the decay of QNM, but the lifetime of the neutron will be zero.

In the SCL, micro-phenomena and cosmic phenomena are largely decoupled with separate dynamics (Hamiltonians) and different degrees of freedom. An implication of the decoupling is that in the SCL, if we work only  in microscopic units we simply cannot tell if we are in de Sitter space  or in flat space. Any common origin of the cosmic and microscopic behaviors only becomes apparent in corrections to the SCL, which typically have the form of powers of the inverse entropy, inverse de Sitter radius, or in the holographic description,  powers of $1/N$.

\bn

In order to flesh out these abstract principles and see the actual machinery at work  a concrete example would be  invaluable.  This is why I've been  interested in double-scaled SYK at infinite temperature (\dk)  as a model of de Sitter  holography. My goal in this paper is  to  explain how the various relevant scales  emerge in  a consistent way  on both sides of the duality. Although far from a proof, this consistency gives some support to the conjecture.

 \subsection{J-T from 3-D De Sitter}

 The specific de Sitter space conjectured to be dual to \dk \ is J-T gravity with a positive cosmological constant \cite{Adel}. The theory is a dimensional reduction of three-dimensional Einstein gravity with a positive cosmological constant. The de Sitter solution is described by the metric,
\be 
ds^2 = -(1-\frac{r^2}{\l^2})dt^2 +\frac{dr^2}{(1-\frac{r^2}{\l^2})} +r^2 d\alpha^2
\label{metric}
\ee
The dimensional reduction is on the $\alpha$ coordinate, leading to a two-dimensional theory with a dilaton field  $\Phi$ equal to $r$. 

\bea
ds^2 \eq  -(1-\frac{r^2}{\l^2})dt^2 +\frac{dr^2}{(1-\frac{r^2}{\l^2})} \cr \cr
\Phi &=& r.
\label{dilaton}
\eea
The cosmological scale is defined by the parameter $\l.$ 
Surprisingly (as we will see) the micro scale in three dimensions is subtle and is not the same as the Planck length.

\begin{figure}[H]
\begin{center}
\includegraphics[scale=.4]{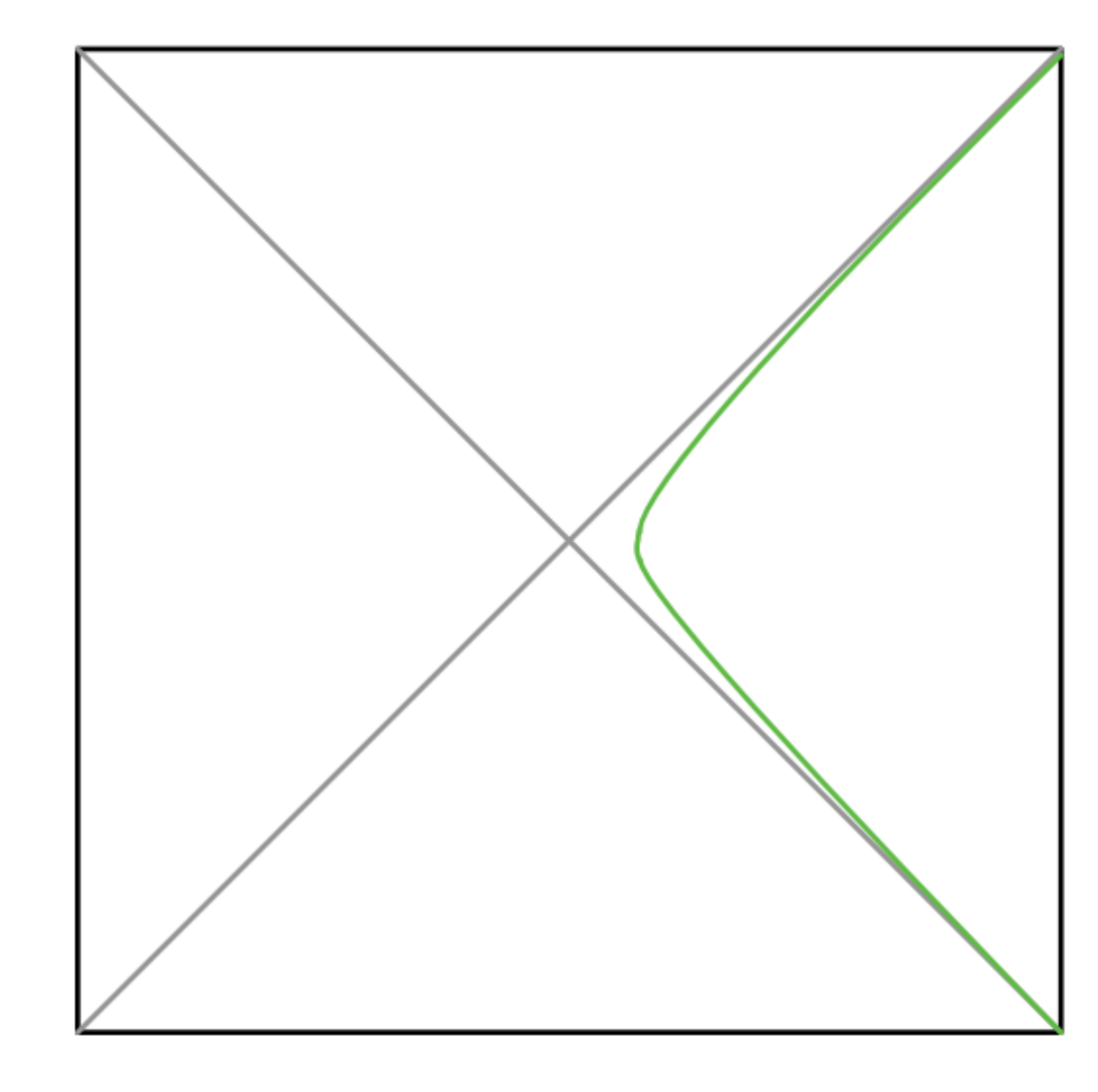}
\caption{Penrose diagram for de Sitter space showing two static patches. The green surface represents the stretched horizon of the right-side static patch.}
\label{penrose}
\end{center}
\end{figure}

The framework for this paper is static patch holography as defined and  described in \cite{Susskind:2021esx}\cite{Susskind:2022dfz} and summarized in \cite{Adel}. The
 holographic degrees of freedom (the hologram) are located on the stretched horizon.
The Penrose diagram in Figure \ref{penrose} shows such a static patch as the right-triangular region. The green curve shows the stretched horizon.
Time-evolution of the holographic \dof \ may be thought of as evolution along the stretched horizon. The static patch itself is a \it bulk \rm reconstruction from the hologram.
\bn

In the  following I will set $c=\hbar =1$ but leave Newton's constant explicit. The four and three dimensional Newton constants are denoted $G_4$ and $G_3.$

\sc
\section{Mass and Length Scales in dS}

What are the limitations on using a flat-space approximation in analyzing phenomena in de Sitter space? There are two; the first limits the amount of mass to be small enough that it does not cause a global back reaction on the geometry.  The other requires the energy to be large enough that the corresponding wavelength is shorter than the de Sitter scale. The first limitation is gravitational and entirely classical; the second involves quantum mechanics.  
The flat space approximation is good in the region sandwiched between these limits. 

\subsection{4-D}

Let us begin with the case of 4-dimensions. 
The maximum mass/energy in 4-D de Sitter space without severe global back reaction is of order,
\be  
M_{max} \sim  \frac{\l}{G_4}  
\label{Mmax4}
\ee
This nominally corresponds to the mass of a black hole with a \S \ radius comparable to the de Sitter radius $\l.$ 

The second limitation is at the opposite end of the energy spectrum: the energy should be large enough so that the corresponding wavelength is significantly shorter than  $\l.$ The minimum energy that satisfies this criterion is
\be
M_{min} = 1/\l. 
\label{Mmin4}
\ee

\begin{figure}[H]
\begin{center}
\includegraphics[scale=.4]{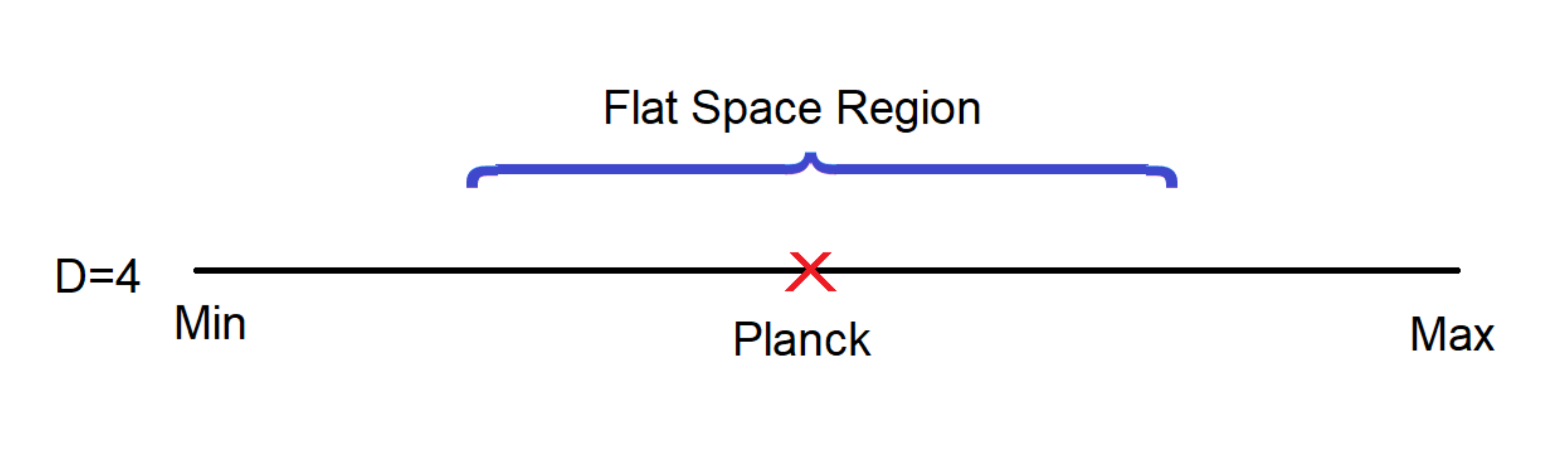}
\caption{Mass  scales in 4  dimensional dS. The horizontal axis represents $\log{M}.$ Mass scales increase to the right and length scales decrease. }
\label{scales4}
\end{center}
\end{figure}

At the extreme ends of the spectrum the curvature of dS cannot be ignored, but for 
$$M_{min}<<M<<M_{max}$$ flat space is a good approximation. The flat space region is centered around the geometric mean of $M_{min}$ and $M_{max}.$ Let's call it 
$M_{m}$ ($m$ denoting microscopic, mean, or middle),

\bea
M_{m} &=& \sqrt{M_{min}M_{max}}  \cr \cr
\eq \sqrt{\frac{1}{G_4}}
\label{=1/G4}
\eea
This happens to be exactly the four-dimensional  Planck mass,
\be 
M_m = M_p    \ \ \ \ \ \ \ \ \ \ (\rm in \ 4 \ dimensions)
\label{Mm=Mp}
\ee
The Planck mass     lies at the center (logarithmically) of the flat-space region, midway between $M_{min}$ and $M_{max}$ as shown in figure \ref{scales4}

The flat space region is very broad in our world, including everything from elementary particles to galaxies. In the semiclassical limit in which the entropy goes to infinity all of these scales are presumed to stay fixed in Planck units.

The two length scales of interest are,
\bea 
\l \eq 1/M_{min} \cr \cr
 L_{m} \eq  L_{p}   \cr 
 \eq 1/M_{m}\cr 
 \eq 1/M_p,
\label{length4}
\eea
the subscript $p$ referring to Planck.

Evidently, in 4-dimensions the Planck mass plays a dual role. As always it controls the density of entropy on horizons through the Bekenstein law, and it defines the mid-point $M_{m}$ lying between the extremes $M_{max}$ and $M_{min}$. These are two quite different concepts, but in 4-D they define the same mass.
\bn

Another scale that will concern us is the string scale $L_s.$ In 4- dimensions it is given in terms of the  string-coupling and the Planck scale,
\be 
L_s = \frac{L_p}{g^2} =\frac{L_m}{g^2}
\label{lstring}
\ee

\bn


\bn

The various length scales, $\l, \ L_m, \ L_s$ allow us to define systems of units. For example time measured in units of $\l$ will be called cosmic time, $t_c$. Similar considerations apply to micro and string time:
\bea
t_c \eq  \lf  \frac{t}{\l} \rg \l    \cr \cr
t_m \eq  \lf  \frac{t}{L_m} \rg  \l   \cr \cr
t_s \eq  \lf  \frac{t}{L_s} \rg  \l
\label{tunits}
\eea
The quantities in the parenthesis represent time measured in cosmic units,  micro units, and  string units. The universal factor of $L_c$ is just for dimensional consistency, in order to give  $t_c, \ t_m,$ and $t_s$  dimensions of length.
(Note that while $\l, \ L_m, \ L_s$ represent definite lengths, the quantities $t_c, \ t_m, \ t_s$ represent the   time variable  measured in different units.)

As examples consider the decay time of quasinormal modes and the decay time of the neutron. The decay time of a QNM in de Sitter space is order-one in cosmic units,  but it  is much larger in micro units where it diverges in the SCL.  
By contrast the decay time of a neutron is parametrically order-one in microscopic units but is very small in cosmic units, vanishing in the semiclassical limit.

\subsection{3-D} \label{3-D}
\begin{figure}[H]
\begin{center}
\includegraphics[scale=.4]{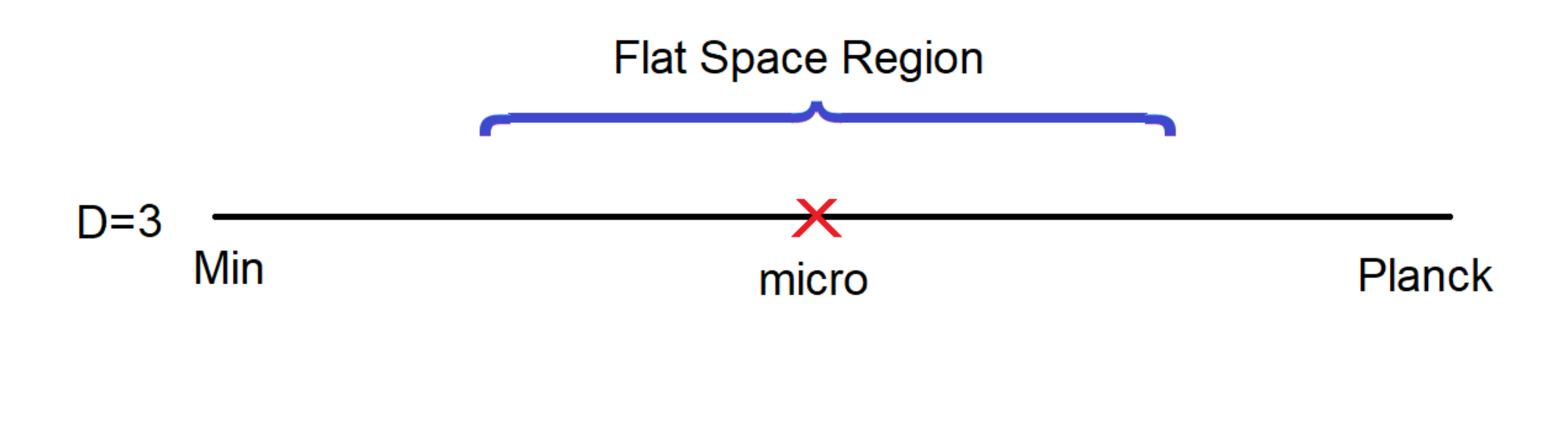}
\caption{Mass  scales in 3  dimensional dS}
\label{scales3}
\end{center}
\end{figure}

From here on we will be concerned only with 3-dimensions where
the situation shown in figure \ref{scales3}  is quite different. 
\bn
To understand figure \ref{scales3} we note that the largest mass that can be placed in 3-D de Sitter is the 3-D planck mass (equal to $1/G_3).$ 
\be 
M_{max} = M_p = \frac{1}{G_3}.
\label{mmax3}
\ee
This is because a mass in three dimensions creates a conical deficit. The Planck mass gives a limiting deficit of $2\pi.$ The back-reaction on the geometry is so strong that the geometry becomes a periodically-identified thin sliver when the mass approaches $M_p.$
  
On the other hand the smallest mass for which the wavelength is  $\geq \l$ is the same as before,
\be
M_{min} = 1/\l. 
\label{mmin3}
\ee
 The middle/mean/micro mass---the geometric mean of $M_{max}$ and $M_{min}$---is,
\bea 
M_{m}&=&   1/L_{m}    \cr \cr
 &=&  \frac{1}{\sqrt{\l}}  \frac{1}{\sqrt{G_3}} .
\label{mm}
\eea
Unlike in 4-D the mass $M_m$ is not the Planck mass.
What was a single scale in 4-D---the Planck mass---generalizes in two different directions in 3-D: one direction being  the usual Planck mass governing the density of horizon entropy;  the other being  the microscopic mass $M_m$ lying midway between $M_{max}$ and $M_{min}$.

In $D=3$ equations \eqref{length4} are replaced by,
\bea 
\l \eq 1/M_{min}  \cr \cr
L_{m} \eq 1/M_{m} = \sqrt{\l} \sqrt{G_3}
\label{length3}
\eea

As in 4-D we can define cosmological and microscopic units of time,
\bea
t_c \eq \lf  \frac{t}{\l} \rg  \l     \cr \cr
t_m \eq \lf \frac{t}{L_{m}} \rg   \l
\label{units3}
\eea
Their ratio satisfies,
\bea
\frac{t_m}{t_c} \eq \sqrt{\frac{\l}{G_3}} \cr \cr
&\sim& \sqrt{S_{ds}},
\label{ratiot3}
\eea
where $S_{ds}$ is the usual de Sitter entropy.
Equation \eqref{ratiot3} shows that the ratio of microscopic and cosmic scales diverges in the semiclassical limit $S_{ds} \to \infty.$

The two time-variables $t_c$ and $t_m$ have associated conjugate Hamiltonians $H_c$ and $H_m$ which scale inversely to their times,
\bea
H_m \eq i\frac{d}{dt_m} \cr \cr
H_c \eq i\frac{d}{dt_c}.
\label{Hs}
\eea
Their ratio  satisfies,
\bea
\frac{H_c}{H_m} \eq 
\sqrt{\frac{\l}{G_3}} \cr \cr
&\sim&   \sqrt{S_{ds}} .
\label{ratioh3}
\eea

\subsection{String Scale}\label{Ss :string}

\dk \ is not in any obvious way a string theory (non-obvious is another matter). Nonetheless  there is 
one additional length or mass scale  in \dk \ which  plays a role similar to the string scale.  I'll call it $L_s$
although it's relation to actual  strings is unclear at the moment. 
 In \dk \  the scale $L_s$  is   dynamical, emerging from SYK dynamics, similar to the  way the string scale emerges from gauge theory dynamics in AdS/CFT.  Its origin is the subject of section \ref{S:strscale}.

\begin{figure}[H]
\begin{center}
\includegraphics[scale=.6]{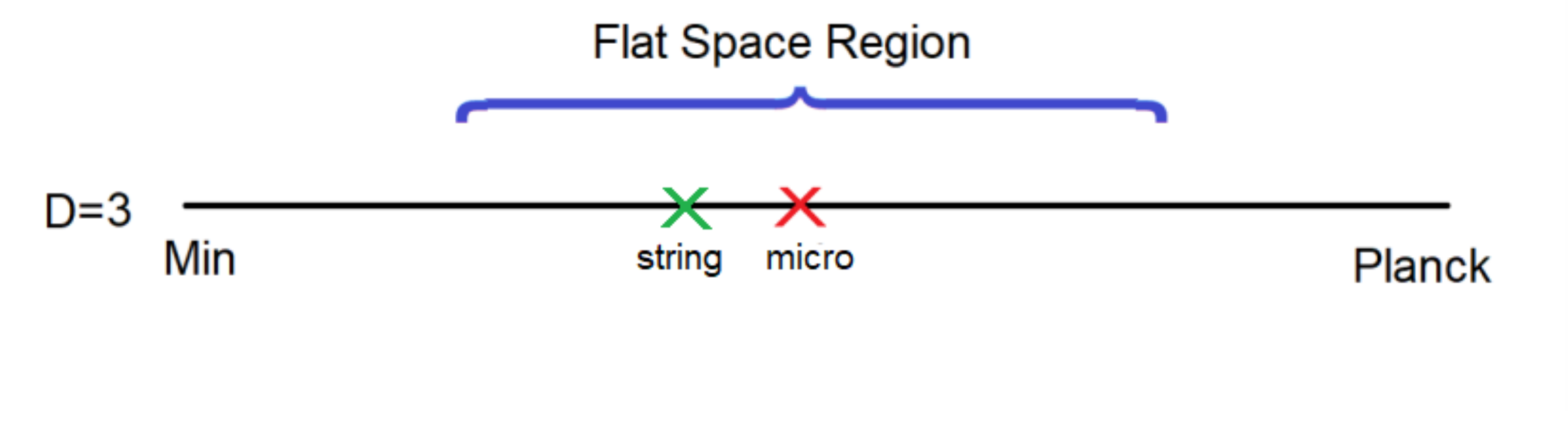}
\caption{Mass  scales in 3  dimensional including the string scale. Typically the string mass scale will lie near but somewhat below the geometric mean of $M_{min}$ and the Planck mass. }
\label{allscales}
\end{center}
\end{figure}

The string scale is microscopic although typically larger than than $L_m$ as shown in figure \ref{allscales}

 We define a parameter $\lambda$ by,
\be 
\lambda = \lf \frac{L_m}{L_s} \rg^2
\label{ls3}
\ee
Equation \eqref{ls3} is the definition of $\lambda$ in terms of gravitational and string quantities. Later we will see that
$\lambda$ also has a dual meaning as a parameter in \dk.

If the parameter $\lambda$ is held fixed at order $1$ then $L_s$ will be of the same order as the microscopic scale $L_{m}$. On the other hand $\lambda$ is a tunable parameter which if taken  very small will lead to a string scale much larger than $L_m.$

The string scale allows us to define string units for both  time and for the Hamiltonian,
\bea 
t_s &=& \lf \frac{t}{L_s} \rg \l \cr \cr
H_s \eq H_c      \lf \frac{L_s}{L_c}    \rg
\label{sunits}
\eea

\sc
\section{DSSYK$_{\bf {\infty}}$} \label{dssyk}

The standard SYK model \cite{Maldacena:2016hyu} is defined as a system of $N$ real fermionic degrees\footnote{The fermion \dof \ are denoted $\chi$ and are normalized so that $\chi^2 = 1.$} of freedom coupled through $q$-local interactions,
\be 
H_{s} = \sum J_{i_1i_2 i_3...i_q}\chi_{i_1}\chi_{i_2}\chi_{i_3}...\chi_{i_q}.
\label{Hsyk}
\ee
The couplings $J$ are drawn from a gaussian ensemble with variance,
\be 
\la JJ\ra =\frac{q!}{q^2N^{q-1}}\CJ^2.
\label{varMS}
\ee
$\CJ$ is a fixed parameter with units of energy.
The SYK model is defined by letting $N\to \infty$ while keeping $q$ fixed.

The \dk \ model  \cite{Cotler:2016fpe}\cite{Berkooz:2018jqr}\cite{Lin:2022rbf}   is similar,  the difference being that we  let $q$  grow with $N$ according to,
\be 
\frac{q^2}{N} = \lambda \ \ \ \ \ \ \ \ \ ( \lambda \ \rm fixed)
\label{q2/N=lamb}
\ee
The parameter $\lambda,$ defined purely in terms of \dk \   in \eqref{q2/N=lamb},  is conjectured to be dual to the parameter  $\lambda$  that appeared in \eqref{ls3}.

We will find that 
the standard Hamiltonian in \eqref{Hsyk} is scaled in such a way that it corresponds to string units.  In the SCL ($N\to \infty$ limit  of \dk) $H_s$ describes evolution in string units.
In other words it is the Hamiltonian defined 
in equation \eqref{sunits},
$$H_s= i \frac{d}{dt_s}.$$

\bn

In \cite{Lin:2022rbf} another Hamiltonian was introduced which we will call $H_{c}.$ It is defined by,
\bea 
H_{c} \eq  \sum j_{i_1i_2 i_3...i_q}\chi_{i_1}\chi_{i_2}\chi_{i_3}...\chi_{i_q}
\cr \cr
\la jj\ra &=& \frac{q!}{N^{q-1}}\CJ^2.
\label{Hc}
\eea
The couplings $J$ and $j$ are simply related,
\be 
j=qJ
\ee
As the notation suggests, $H_c$ is the Hamiltonian in cosmic units,
\be 
H_c = i\frac{d}{dt_c}.
\ee
For finite values of $N$ and $q$ the two Hamiltonians $H_s$ and $H_c$  differ but only  by a factor of $q$,
\be 
H_c = q H_s.
\label{Hc=qHs}
\ee
In the SCL $q\to \infty$ and the ratio of $H_c$ and $H_s$ diverges.

Finally, for reasons explained in \cite{Lin:2022rbf}, to define \dk \  the temperature (defined through the Boltzmann distribution) is taken to infinity  
\be 
T= \infty.
\ee
As a consequence the density matrix of the static patch is maximally mixed as advocated in  \cite{Chandrasekaran:2022cip}.

\subsection{Perturbation Theory}
In this subsection I will review some facts about  SYK perturbation theory at infinite temperature,  which I assume the reader is generally familiar with \cite{Roberts:2018mnp}. The expansion parameter is $\CJ.$ The only unusual thing is that we will work with the Hamiltonian in \eqref{Hc} in cosmic units
whereas the usual analysis is in string units. 

\begin{figure}[H]
\begin{center}
\includegraphics[scale=.5]{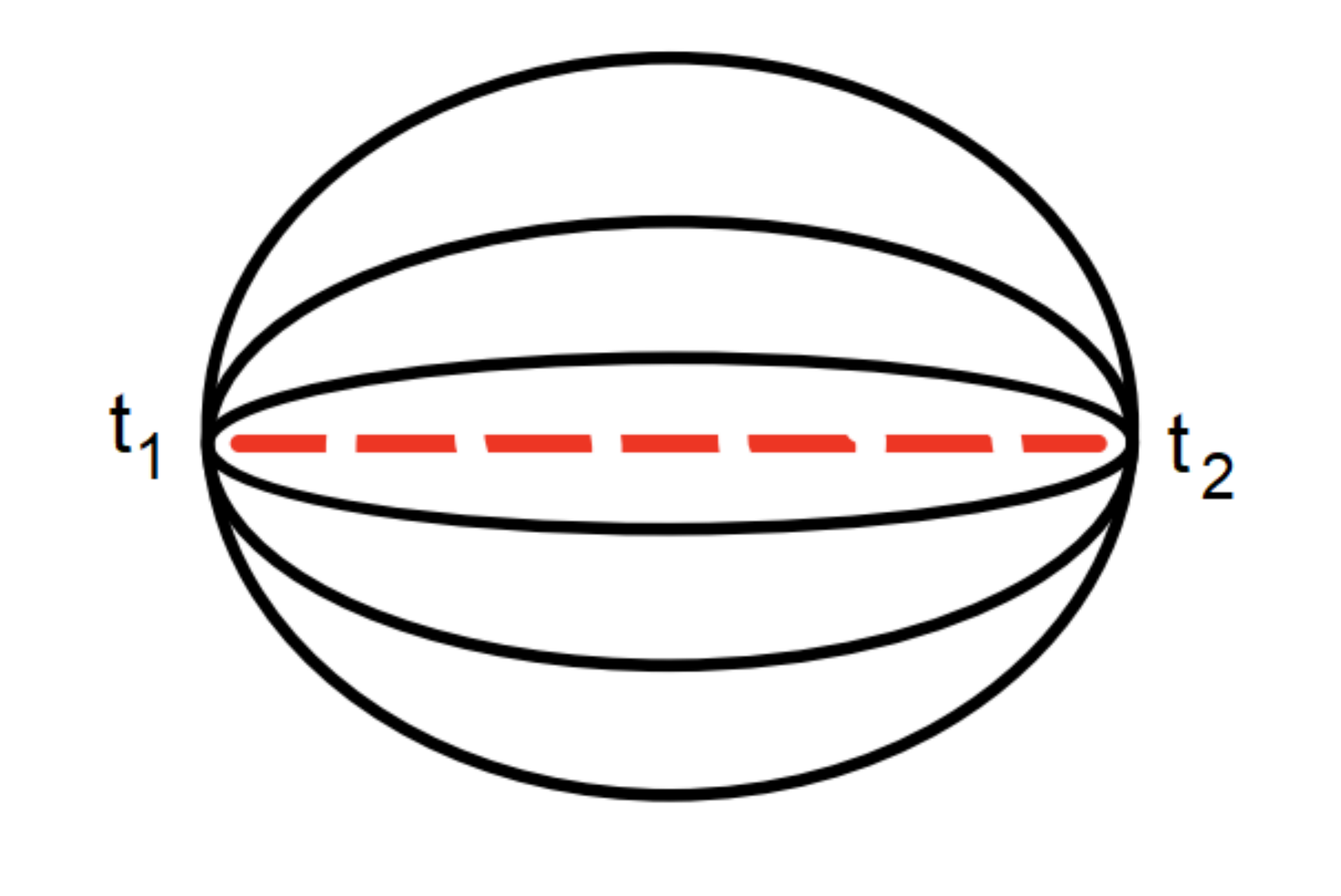}
\caption{The simplest vacuum melon diagram.}
\label{melon}
\end{center}
\end{figure}

The vertices of the Feynman diagrams have numerical value  $\CJ.$ The fermionic propagators are represented by solid black lines. At infinite temperature they are very simple, given by  $$\epsilon(t_2-t_1) = \text{sign}(t_2-t_1)$$ where $t$  refers to cosmic time.
Ensemble averages (Equation \eqref{Hc}) over the gaussian probabilities for 
$j$ are represented by broken red lines.
The discussion will be very brief and will focus on a few specific diagrams as examples.
To illustrate let's consider the vacuum diagram in figure \ref{melon}.
\bn
For definiteness I've drawn the diagram for the case $q=6$ but the generalization is obvious.
Since there are an even number of propagators  in figure \ref{melon}   the overall sign is positive. 

The value of the diagram is
\be 
\text{vac diag } = \CJ^2   \     \frac{N^q}{q!}   \times    \frac{q!}{N^{q-1}}
\label{vacdiag}
\ee
The first factor after $\CJ^2$ is   the number of ways of choosing $q$ fermion modes from a total of $N$,
$$\frac{N!}{q!(N-q)!} \approx \frac{N^q}{q!}.$$
The second factor is the
 the correlator  $\la jj\ra$  in \eqref{Hc}.  
  The  result is 
\be
\text{vac diag } = \CJ^2  N.
\label{vacdiag2}
\ee
If for some reason one wants to integrate the diagram over the relative time between vertices the expression would be,
\be
\text{vac  diag } = \CJ^2  N \int dt,
\label{vacdiag2}
\ee
The result  would be infrared divergent because the integrand is independent of $t$.

\begin{figure}[H]
\begin{center}
\includegraphics[scale=.4]{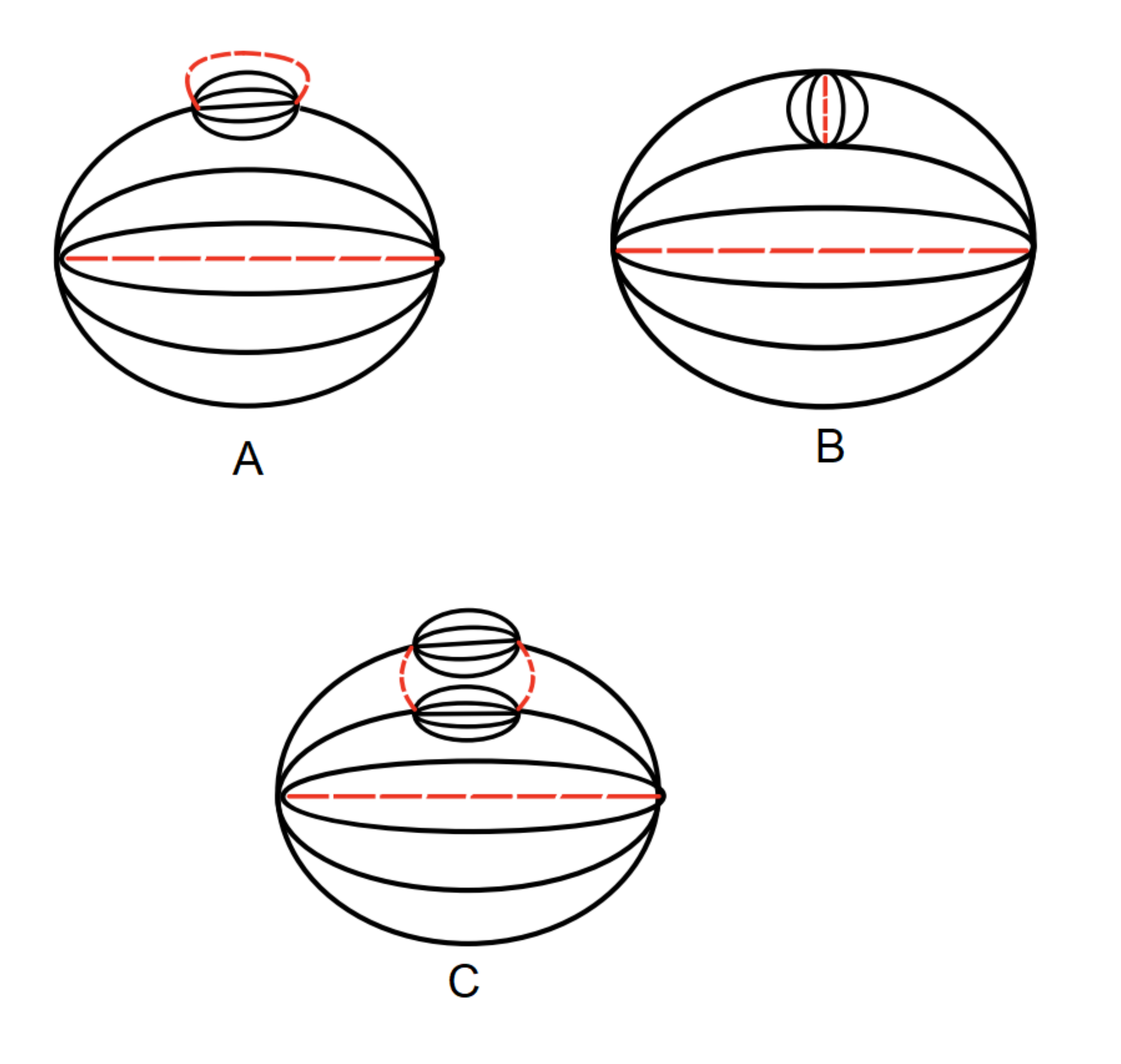}
\caption{Additional vacuum diagrams.  \ (A) another melon diagram. \  (B) a non-melonic diagram to next order in $1/N$. \ (C) a non-perturbative diagram in the $1/N$ expansion.}
\label{diagrams}
\end{center}
\end{figure}

In figure \ref{diagrams} three more typical diagrams are shown.
 Diagram $\bf A$ is another melon diagram whose leading behavior is,
\be
\CJ^4 q \frac{N^q}{q!} \frac{N^{q-1}}{(q-1)!} \times  \frac{q!}{N^{q-1}}  \frac{q!}{N^{q-1}} \int d^3 t,
\label{A}
\ee
where the integral is an appropriate three dimensional integral.  I will not spell it out further except to say that it is IR-divergent. All together we get,
\be 
\CJ^4 N q^2 \int d^3 t
\label{vacdiag3}
\ee
The expression \eqref{vacdiag3} scales the same way with $N $ as \ref{vacdiag2} but contains an extra factor of $ q^2$. 

The next example,  diagram   \ref{diagrams} $\bf{B},$  is non-melonic. It has the value,
\bea 
&& \CJ^4 q(q-1) \frac{N^q}{q!}     \frac{N^{q-2}}{(q-2)!} \times   \frac{q!}{N^{q-1}}     \frac{q!}{N^{q-1}}   \int d^3 t  \cr \cr
\eq \CJ^4 q^2(q-1)^2   \int d^3 t 
\label{vac4}
\eea
This non-melonic diagram is smaller by a factor of $N$ than the melonic diagrams, but  higher order in $q$ than the leading melon diagram. 

Next let us look at figure \ref{diagrams}-$\bf{C}$. This diagram is different than the previous ones in that the dotted red lines---the $jj$ correlators---are not contained within melonic structures. This is a signal that the diagram is non-perturbative in the $1/N$ expansion. 
It is given by,
\be  
\CJ^4   (q-1)q\frac{N^q}{q!}    \frac{N^{q-2}}{(q-2)!}  \times   \frac{q!}{N^{q-1}}    \frac{q!}{N^{q-1}}        \frac{q!}{N^{q-1}}   \int d^5 t  
\ee
Notice that there are more factors of $N^q$ in the denominator than the numerator. To leading order in $q$ the final result is,
\be
\sim  \CJ^4 \lambda^2 N^4  \lf  \sqrt{\frac{\lambda}{N}}   \rg^{\sqrt{\lambda N}}
 \int d^5 t 
\ee
This nonperturbative contribution (NPC) vanishes exponentially faster than any power of  $N^{-1}$ in the double-scaled limit. It is similar to instanton contributions to the large $N_{ym}$ expansion of gauge theory amplitudes, and rapidly vanishes for large $N$.

These examples suggest (correctly) that the general expansion has the form
\be 
\text{amplitude}= N^a \sum_{n=0}^{\infty}\frac{P_n(q)}{N^n} + NPC
\label{dkexpansion}
\ee
The value of $a$ depends on the particular amplitude, for example $a=1$ for the vacuum amplitude. 
   $P_n(q)$ is an infinite order polynomial in $q,$ with the order of the first term increasing with $n$. This is  analogous to the  large $N_{ym}$ expansion of gauge theories with $q$ playing the role of the  't Hooft coupling constant. This observation adds an interesting twist to the \dk \ formula  
$$ {q^2 } = \lambda  N.$$ It  parallels 't Hooft's definition
\be  
\alpha = g^2 N_{ym},
\ee
suggesting  that  $\lambda $ plays a role  similar to the gauge coupling $g^2$, and that $q$ is like the 't Hooft coupling $\alpha$. 

\sc
\section{The Emergent String Scale} \label{S:strscale}

In large $N_{ym}$ QCD there is an emergent scale that is not visible in  perturbation theory, but which  could be seen if it were possible to  sum all planar diagrams. Various manifestations  of it exist: confinement;  the energy gap; the string scale; Regge trajectories; the Hagedorn temperature;   and so on. One of the roles of the emergent scale is to regulate the infrared divergences that appear in perturbation theory due to the masslessness of the gauge bosons.

To see that there is an emergent dynamical scale in \dk \ let us return to figure \ref{melon} and the infrared divergent expression \eqref{vacdiag2}. 
The diagram contains a numerical factor and an integral over the relative cosmic times of the two vertices.
 The bare propagators are $\epsilon(t_2-t_1)$ and when combined give an integrand which is independent of the relative time, thus leading to an IR divergence.
 
\begin{figure}[H]
\begin{center}
\includegraphics[scale=.4]{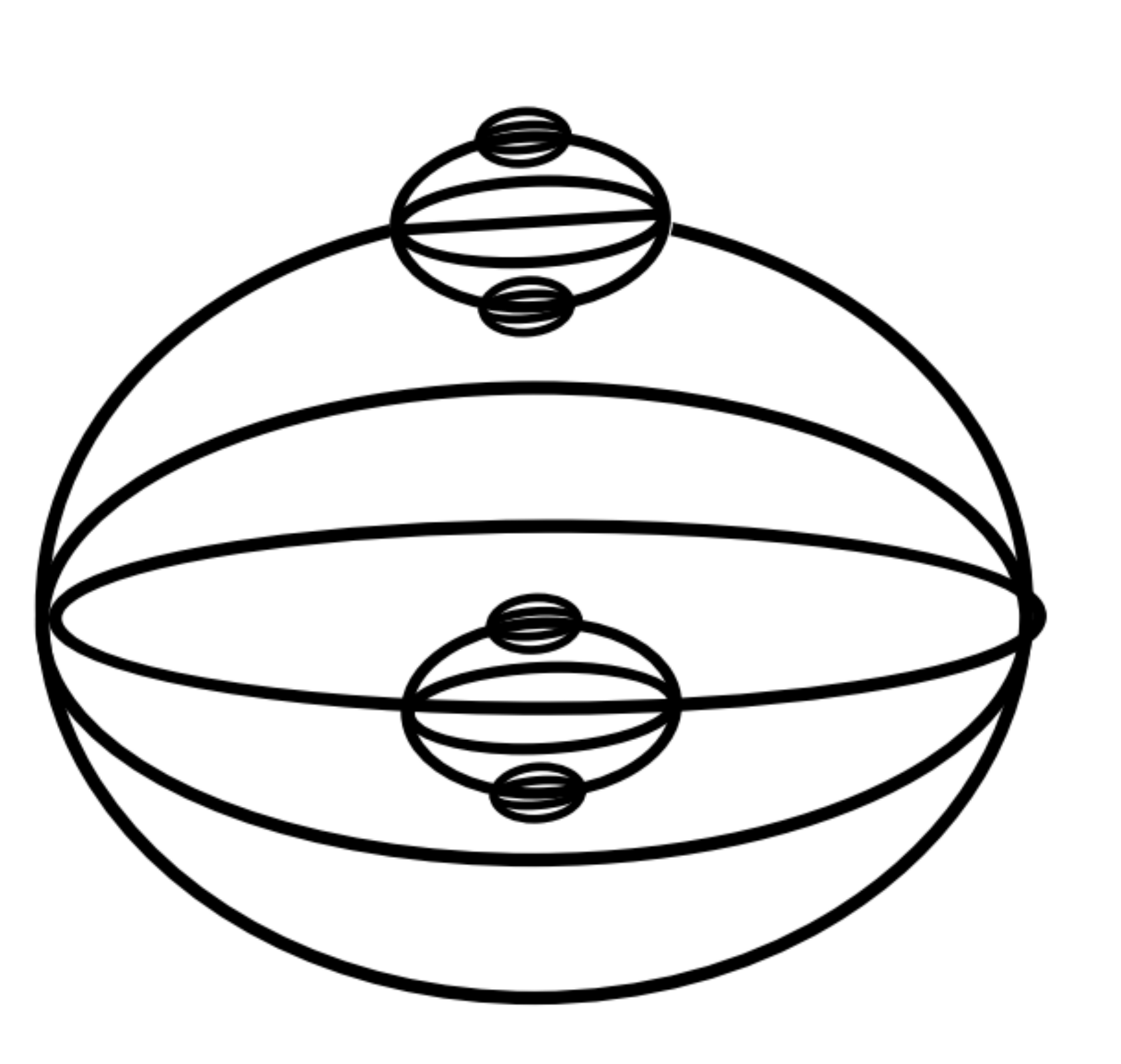}
\caption{An infinite class of diagrams that can be summed by solving the Schwinger-Dyson equation.}
\label{bubbles}
\end{center}
\end{figure}

 However, figure \ref{melon} is just the first of an infinite number of melonic diagrams  in which the propagators are corrected by additional melons,  melons within melons, ad infinitum as illustrated in figure \ref{bubbles}.
 Each of these diagrams is IR divergent but they can be formally summed for small $\lambda$ by solving the Schwinger-Dyson equations  \cite{Maldacena:2016hyu}. 
The result  is that the  trivial  integrand is replaced by the nontrivial function \cite{Roberts:2018mnp},
\be  
\frac{1}{\cosh^2{\CJ q |t_1-t_2|}} = \frac{1}{\cosh^2{\frac{|t_1-t_2|}{L_s} }}
\label{cosh2}
\ee
Two things are evident. The first is that a new  scale $L_s$  appears. Defined  by
\be 
L_s =\frac{1}{\CJ q},
\label{Ls=1/Jq}
\ee
the new scale $L_s$ is an emergent non-perturbative manifestation of the SYK dynamics.

Secondly, the new scale regulates the infrared divergences. For example, the divergent integral 
in \eqref{vacdiag2} is replaced by,
\be 
\int dt \to \int dt  \frac{1}{\cosh^2{\CJ q t}} \sim \frac{1}{\CJ q}
\ee
More generally integrals like $$\int d^nt$$ are replaced by $$c \lf \frac{1}{\CJ q} \rg^n$$ with $c$ being a diagram-dependent constant of order unity.

\sc
\section{  Tomperature and the Cosmic Scale}
The simplest way to make contact between \dss \ and \dk \ is to relate \dk \ quantities to the de Sitter length scale $\l.$ This brings us to the concept of ``tomperature"   \cite{Lin:2022nss}
.
The Boltzmann  temperature of the \dk \ model is infinite, but that does not mean that the ``effective" temperature of the radiation in the static patch is infinite. The Gibbons-Hawking temperature is not the Boltzmann temperature $T$ but rather the tomperature $\CT$ \cite{Lin:2022nss}. 

Hawking radiation is a cosmic phenomenon. The wavelength of the radiation, the energy of a Hawking quantum, and the rate of emission are all order $1$ in cosmic units, so it is only reasonable to describe it  in cosmic units. Indeed 
Tomperature is defined as the change in \it cosmic \rm energy \rm  $E_c$ (expectation value of the cosmic Hamiltonian $H_c$) if one qubit (two fermion \dof) is removed,  keeping fixed all those couplings not involving the deleted  qubit.  It can be understood in terms of a version of the first law,
\be 
\Delta E_c = \CT \Delta S.
\ee
 It was shown in \cite{Lin:2022nss} that the tomperature is given by,
\be 
\CT = 2\CJ
\label{T=J}
\ee
Since the observed temperature in the de Sitter static  patch is $1/(2\pi \l)$ it follows that 
\be
\l \sim 1/\CJ.
\label{lc=1/J}
\ee
This establishes a point of contact between gravitational de Sitter quantities and \dk \ quantities that will allow us to build the start of a dictionary in the next section.

 \bn
 
From equations \eqref{Ls=1/Jq} of the last section, and \eqref{lc=1/J} of this section,  we clearly see that in the SCL  a  separation of scales, separated  by a factor of $q,$ takes place in \dk.
This separation  between the cosmic and string scales parallels a corresponding separation that takes place in de Sitter space. 

\sc
\section{DSSYK- De Sitter Correspondences}\label{ds-dssyk}
Equation \eqref{lc=1/J} provides a link connecting \dk \ parameters to de Sitter parameters.
Here are some correspondences that follow.
 \begin{enumerate}

 \item 
 
  From \eqref{T=J} and \eqref{lc=1/J},
 
 \bea 
 \l &\sim& 1/\CT    \cr \cr
 &\sim& 1/\CJ
 \eea

 \item
 \be  
 S_{ds} \sim N
 \ee
This follows from the assumption of infinite temperature: The entropy is equal to the number of degrees of freedom.
 
 \item
\be 
S_{ds}\sim \frac{\l}{G_3} \sim \frac{\l}{L_p}
\label{bek}
\ee 
Equation \eqref{bek} is the  Bekenstein law in 3-D.

It follows that
\be 
L_p \sim \frac{1}{\CJ N}
\ee

and

\be 
\frac{L_p}{\l} = \frac{1}{N}
\ee
 
\item

Now recall that $L_{m} = \sqrt{\l G}$ giving,
\be 
L_{m} = \frac{1}{\CJ \sqrt{N}}.
\label{Lm=1/JsqrtN}
\ee
Using \eqref{Ls=1/Jq} we find,
\be 
\lf  \frac{L_s}{L_m}  \rg^2 =  \frac{1}{\lambda} 
\ee
 \end{enumerate}

The fact that the string and cosmic scales differ by a factor of $q,$
\be
L_s = \frac{1}{\CJ q} = \frac{\l}{q},
\label{factq}
\ee
 justifies our  identifying  $qH_s= H_c$ as in \eqref{Hc=qHs}. 

Note that for $\lambda \sim 1$ the string-scale $L_s$ and the micro-scale $L_m$ are approximately equal---thus justifying the claim that the string scale is microscopic. However, by allowing $\lambda<<1$ the string and micro scales can be separated so that the string length scale is much larger than the microscale. This is entirely analogous to what happens in AdS/CFT where if the gauge coupling is order $1$ the string and Planck scale are approximately the same, but by making 
$g_{ym}^2<<1$ the string scale can be made much larger than the Planck scale.

(Note the similarity with \eqref{lstring} with $\lambda$ playing the role of $g^4$.)

\sc
\section{Gauge and DSSYK$_{\infty}$  Correpondences}   \label{g-dssyk}

I will digress here in order to further discuss the similarity between  \dk \ and the large $N_{ym}$ limit of gauge theories. 
There is a close similarity between perturbation theory in \dk \ and in gauge
theory. We will not go into detail but just remind the reader of some aspects of
   the two perturbation expansions. For illustration we can consider the connected vacuum
   amplitude $V$. It is an infinite sum of the form,
\be  
V = N_{ym}^2 \sum_n \frac{P_n(\alpha)}{N_{ym}^n} +NPC
\label{hooft}
\ee
where $P_n$ is a polynomial (generally infinite order) in $\alpha$. Each term represents the sum of diagrams with the same topology and $n$ is the genus of that topology. In addition there are non-perturbative contributions---for example from instantons---which vanish exponentially in the large $N$ limit.
This $1/N_{ym}$ expansion is very well known and we assume the reader is thoroughly familiar with it.

Next consider the expansion of the SYK vacuum amplitude. It also has a $1/N$ expansion (where $N$ refers to the SYK parameter denoting the number of fermion species) which takes a form very similar to  \eqref{hooft},
\be  
V = N \sum_n \frac{P_n(q)}{N^n} +NPC .
\label{hooft}
\ee
Instead of the sum of planar diagrams the first term in \eqref{hooft} is the sum of all melon diagrams. 

The similarity  between gauge theory and \dk \ require us to make the following correspondences between the two.
\bea 
N_{ym}^2 &\leftrightarrow& N \cr \cr
\alpha  &\leftrightarrow& q   \cr \cr
g^4 &\leftrightarrow& \lambda
\label{ym-syk}
\eea
The defining formula for \dk \ 
$$\frac{q^2}{N} = \lambda$$
parallels the gauge theory formula,
\be 
\frac{\alpha}{N_{ym}} = g^2.
\ee

\subsection {Gauge Theory  and DSSYK$_{\infty}$   Scaling}

Let us now write the basic AdS/CFT relations between parameters and then compare them  with the corresponding dS/\dk \ relations. 
   
\bn
\textbf{Gauge Theory}
\begin{enumerate}
\item 
\be
 \lf \frac{L_p}{L_{ads}} \rm\rg^4   = \frac{1}{N_{ym} }  
 \label{Lp/Lads}
\ee
\item
\be 
\lf  \frac{L_s}{L_{ads}} \rg^4 = \frac{1}{g^2 N_{ym}}
  \label{Ls/Lads}
   \ee
\item
\be
\lf  \frac{L_p}{L_s}   \rg^4 = g_{ym}^2
 \label{Lp/Ls}
\ee
\end{enumerate}
\bn
On the \dk side we have similar relations.

\bn
\textbf{DSSYK$_{\infty}$ }
\bn
\begin{enumerate}

\item
\be
 \lf \frac{L_m}{\l} \rg^2 = \frac{1}{N}
  \label{Lm/Lc}
\ee
\item
\be
\lf \frac{L_s}{\l} \rg^2 = \frac{1}{\lambda N}
\label{Ls/Lc}
\ee
\item
\be
 \lf \frac{L_m}{L_s} \rg^2  = \lambda
 \label{Lm/Ls} 
 \ee
\end{enumerate}

\bn
The relations \eqref{Lp/Lads}\eqref{Ls/Lads}\eqref{Lp/Ls}
 are not the same as \eqref{Lm/Lc}\eqref{Ls/Lc}\eqref{Lm/Ls}. This is not surprising since relations like \eqref{Lp/Lads}\eqref{Ls/Lads}\eqref{Lp/Ls} are not universal. Indeed the exponent $4$ on the left side changes as the dimension changes. 
  But they do have a qualitative similarity once we make the replacements $L_{ads} \to \l$; 
  $L_p \to L_m$;  $N_{ym} \to \sqrt{N}$;  $\lambda \to g_{ym}^4$; and the 
 exponent $4$ by the exponent $2$. 
 
 The reasons for these replacements have been explained but I will repeat them here:
 
 First of all we obviously should replace the AdS length scale by the dS scale 
 $\l$. Both of them represent the length scale of the cosmic geometry---AdS in one case and dS in the other.
 
 Second, the Planck length plays a dual role. In its role as the linear dimension of a qubit of information is not really a length at all,  but rather it defines an area $L_p^{D-2}$ in higher dimension. Its role as a length in 4-dimensions is as the microscopic length $L_m = \sqrt{L_{min} L_{max}}$. In this role it generalizes to 3-dimensions. 
 
 Third $N$ and $N_{ym}^2$ are both the number of degrees of freedom. It's just a historical accident that they are denoted differently.
 
Finally, the exponent $4$ is a special case applicable to five bulk dimensions. There is no reason for it to generalize to the current context.

\subsection{Note on L$_m$}
So far  the scales $\l$ and $L_s$ have the primary roles with $L_p$ and $L_m$ being less prominent. About $L_p$ its role is not so much as a length,  but as the determining factor in the relation between horizon area and entropy. But what of $L_m$; does it have any significant importance for the dynamics of \dk?

To see the meaning of $L_m$ lets return to \eqref{ls3},
$$\lambda = \lf  \frac{L_m}{L_s}  \rg^2$$
Since $L_s$ and the parameter $\lambda$ have  physical significance, so must $L_m$.
When $\lambda$ decreases the string scale increases in the same way as it does in AdS/CFT when the gauge coupling decreases. The theory becomes progressively more non-local as $\lambda\to 0.$ So we may say that  the ratio $L_m/L_s$ is a measure of the degree of locality of the holographic bulk, i.e., the static patch.

\bn

\subsection{DSSYK  Is Like the Flat Space Limit of AdS/CFT}
The 't Hooft limit of a gauge theory is the limit of large $N_{ym}$ with $\alpha = g^2N_{ym}$ kept fixed. To make contact with gravity in AdS/CFT  the value of $\alpha$ should be large. From \eqref{ym-syk} one sees that this limit is analogous to the limit $N\to \infty$ with fixed but large $q$ in SYK. This however  is not the limit of \dk. The limit of \dk \ is the limit of fixed $\lambda,$ which corresponds to the gauge theory limit of fixed $g^2.$ This limit is often called the ``flat space limit" \cite{Susskind:1998vk}\cite{Polchinski:1999ry}. The flat space limit is a limit of extremely strong coupling and is notoriously  difficult to control.

Like the SCL of de Sitter space, the flat space limit of AdS also involves a separation  between  micro scales and a cosmic scale, the difference being that the cosmological scale is the scale of AdS rather than de Sitter space. If one wants to study micro physics in the flat space limit, string-scale units are appropriate, but the global AdS nature of the geometry will be lost. If one wants to study the global geometry then cosmic AdS units are more appropriate but local physics will ``shrink" to infinitesimal size.

\sc
\section{Scrambling}

Now let us return to the main subject of this paper, the separation of scales in de Sitter space and the dual separation of scales  in \dk.
Scrambling in \dss \ gives a good example of the  importance of having more than one system of units. 
 In \dk \  scrambling can be described
  \cite{Susskind:2021esx}\cite{Susskind:2022dfz}\cite{Lin:2022nss} by an ``epidemic" probability function\footnote{In previous papers \cite{Susskind:2021esx}\cite{Susskind:2022dfz}\cite{Lin:2022nss}, the cosmic time was denoted by $\t$. In this paper I have replaced the notation $\t$ by  $t_c.$} 
\be 
P(t_c)=1-\lf 1+\frac{q}{N}e^{(q-1)\CJ t_c} \rg^{\frac{-1}{q-1}}
\ee

For small $t_c$ this behaves like,
 \be 
P(t_c) = \frac{e^{(q -1  )\CJ t_c}}{N} 
\label{toofast}
\ee
In the double-scaled limit $q$ becomes infinite, leading to an infinitely rapid rate of increase  \eqref{toofast} in cosmic units. 
But recall the relation between string and cosmic time,
 $$t_s = q t_c.$$
We may write \eqref{toofast} in the form,
\be 
P(t_c)  = \frac{e^{ \CJ  t_s}}{N}
\label{fstscr}
\ee
By switching from cosmic to string units we see a perfectly sensible\footnote{Technically, what I will mean by a ``sensible" formula is one in which neither $N$ nor $q$ appears except in the finite combination $\lambda = q^2/N$. In a sensible formula the phenomenon and the units have been matched in a way that eliminates divergent expressions in the $N\to \infty$ limit.}
exponential operator growth,  characteristic of fast scrambling. 
However, this  behavior only lasts a short time until $P\sim q/N.$  By contrast standard fast scrambling lasts until $P\sim 1.$

In terms of time duration, in string units the fast scrambling in \eqref{fstscr} lasts for a time,
\be 
 \Delta t_s =   \frac{ \log{q}}{\CJ}.
\label{Dts=logq}
\ee
Expressed in terms of cosmic time, as $N$ and $q$ go to infinity the fast scrambling  lasts for,
\be 
 \Delta t_c = \frac{ \log{q}}{q\CJ}. \to 0.
\label{Dtc=logq/q}
\ee
Thus in cosmic time the duration of fast scrambling shrinks to zero in the double-scaled limit. To resolve the early onset of scrambling one must work in  string units.

Scrambling does not stop at the end of this initial period but continues with an entirely different behavior until it saturates at $P=1.$ The transition from one behavior to the other is very sharp in cosmic time.
After the transition the scrambling function $P$ evolves as follows:
\be 
1-P(t_c) = e^{-\CJ t_c},
\label{latetime}
\ee
  lasting for a cosmic time of order $1/\CJ$ 
  until $P\to 1.$  This  ``hyperfast" scrambling (hyperfast in cosmic units) lasts until scrambling is complete, i.e.,   until $P\to 1.$ 

We can  express this late-time behavior in string units,
\be 
1-P(t_s) = e^{-\CJ t_s/q}
\label{latetime2}
\ee
but it would not be sensible; $P$ would  evolve infinitely slowly as $q\to \infty.$  Whether the behavior is hyperfast or extremely slow depends on whether it is viewed in cosmic or micro units.

\bn

To sum up:
the early time scrambling is sensibly described in string units as a period of  fast scrambling, but in cosmic units the early period  shrinks to zero. On the other hand the late time scrambling is sensibly described in cosmic units where it follows the decay law of the leading quasinormal mode.  But if we try to describe the QNM decay in string units we find an infinitely slow evolution once $P(t_s) $ passes the value $q/N.$  To describe the full scrambling behavior in the semiclassical limit, with its very sharp transition, one needs  string units at early time, and cosmic units at late time.

\sc
\section{Correlation Functions}
This section follows work in preparation by Adel Rahman \cite{Adel}.

The holographic \dof \ dual to the static patch are localized on the stretched horizon but the static patch itself is emergent in the sense that it is constructed from the stretched horizon degrees of freedom. In \dk  \ the holographic \dof \ are the fermions $\chi_i$ out of which the static patch emerges through some holographic  dictionary. Let us assume that among the emergent \dof \ there is a light bulk field $f$ that propagates in the static patch. Correlation functions of $f$ could be used to probe the geometry of the static patch. If we can translate those correlations into correlations among the fermionic \dof \ this would take us a long way toward establishing or refuting the conjectured duality between \dk \ and de Sitter space.

\begin{figure}[H]
\begin{center}
\includegraphics[scale=.6]{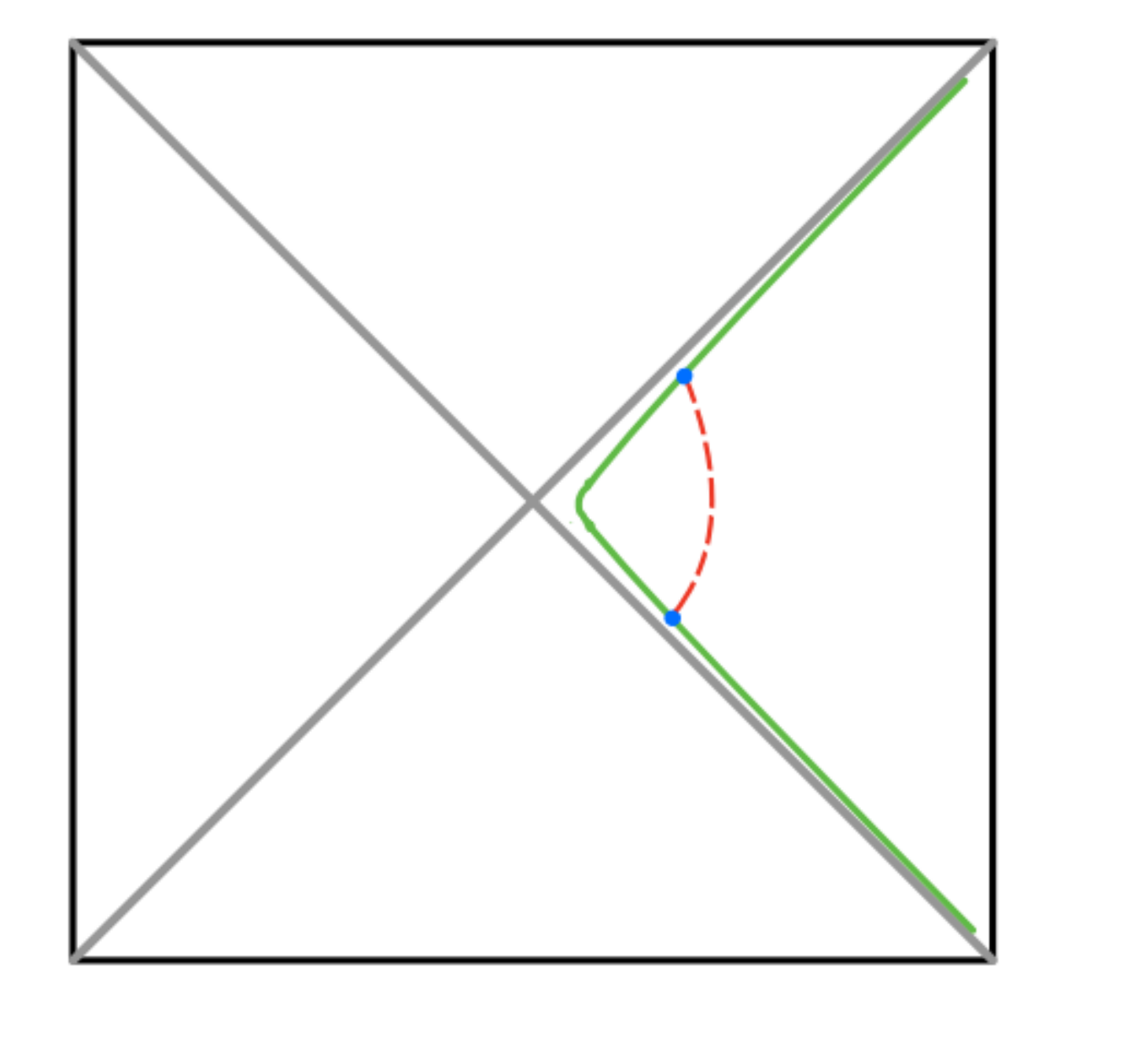}
\caption{Correlation function between two points on the stretched horizon with a relatively small time separation.}
\label{corr}
\end{center}
\end{figure}

\begin{figure}[H]
\begin{center}
\includegraphics[scale=.6]{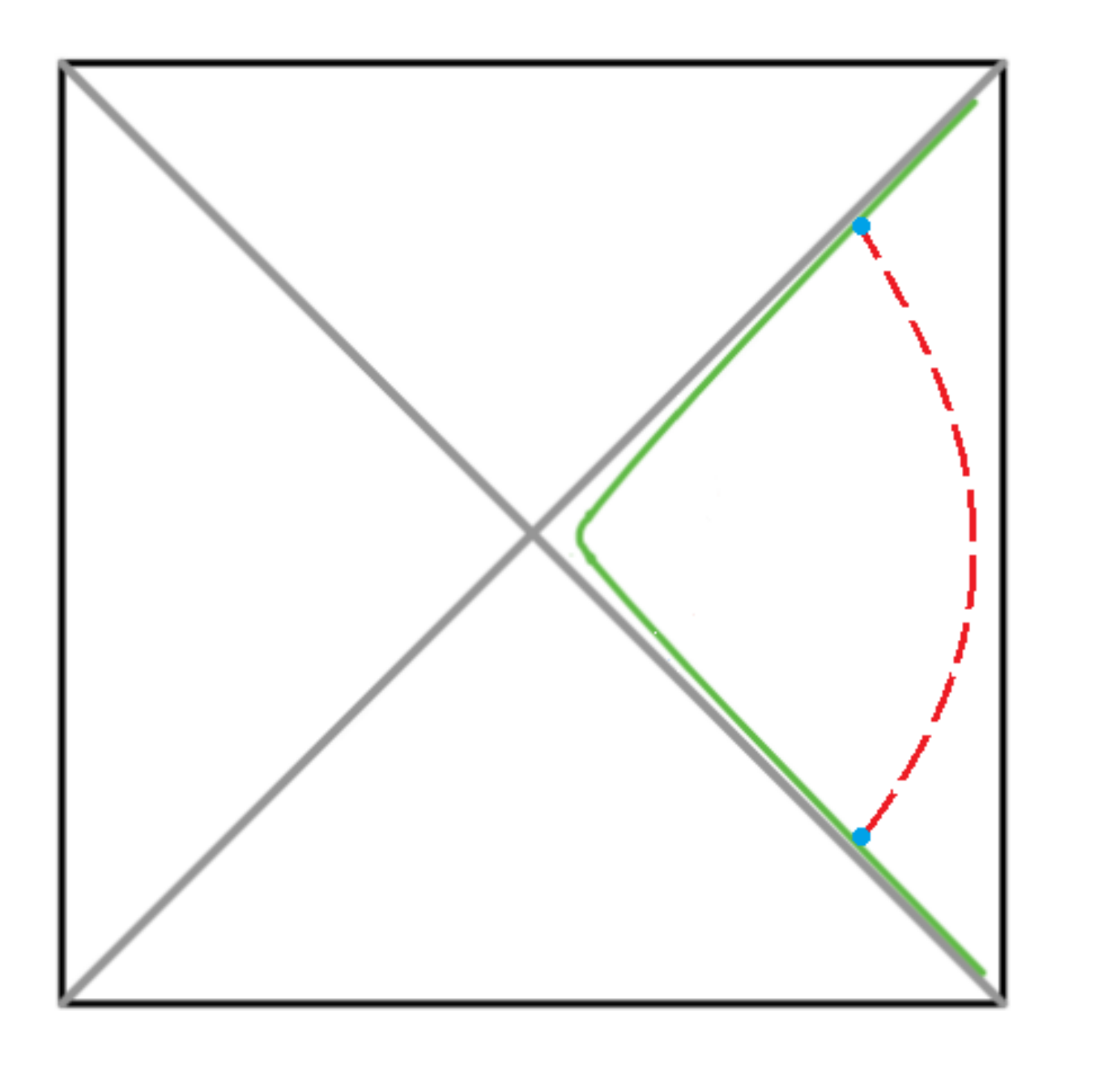}
\caption{Correlation function between two points on the stretched horizon with a relatively large time separation. The larger the time separation the deeper into the static patch the correlations probe.}
\label{deeprobe}
\end{center}
\end{figure}

An example is shown in figure \ref{corr}.
\bn
For a given time-separation the correlation function will be sensitive to the geometry out to some distance from the horizon. 
The larger the time-separation the deeper into the static patch the correlation function will probe \cite{Adel}. The time separation in figure \ref{corr} is relatively small---the correlation probes  relatively close to the horizon. In figure \ref{deeprobe} the time separation is larger and the correlation function is more sensitive to the geometry deep into the static patch.
It is not hard to estimate the required time separation to probe all the way to the center of the static patch (the pode) \cite{Adel}. For a massless particle it takes a time of order $t_* =\l \log{S}$ to reach the pode and return to the horizon. This is the time-scale on which correlations probe deeply into the static patch.

\begin{figure}[H]
\begin{center}
\includegraphics[scale=.5]{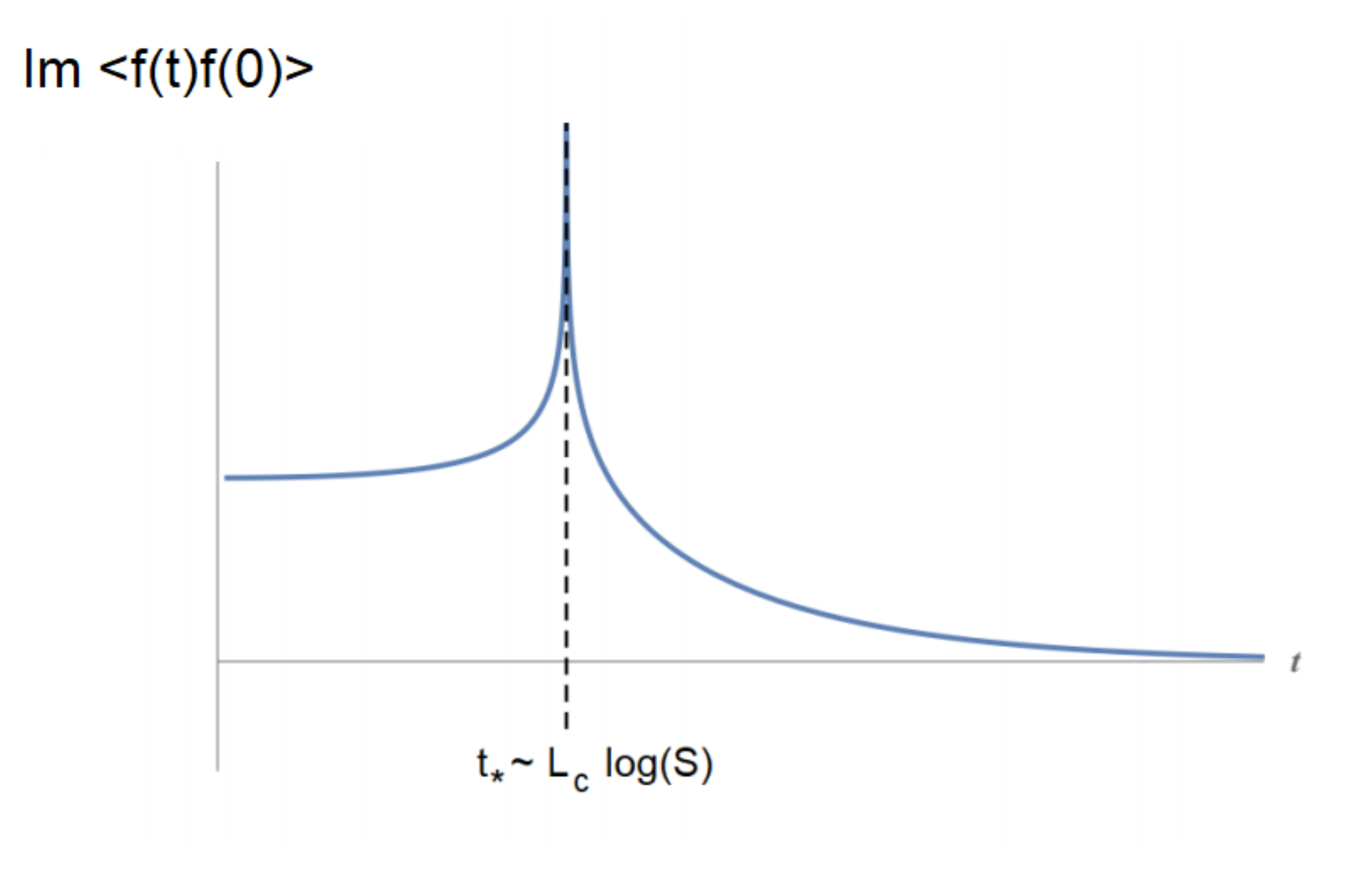}
\caption{Imaginary part of light bulk scalar correlation function calculated by A. Rhaman.}
\label{bump}
\end{center}
\end{figure}

Bulk correlations for light scalar fields coupled to the dilaton, in JT gravity,  have been computed in \cite{Adel}. Figure \ref{bump} shows the imaginary part of correlation function of a scalar field as a function of cosmic time.
Indeed $\la f(0)f(t)\ra$ has a distinct feature at time $\sim \l \log{S}.$
The properties of  $\la f(0)f(t)\ra$ include the following:

\begin{enumerate}

\item
The imaginary part of the correlation is singular at $t_c = t_*$. The exact nature of the singularity will not be important for us, but its origin is  the on-shell propagation of a particle from the horizon to the pode and back.

\item
The correlation persists over  a time $\sim t_* \sim L_c \log(S_{ds}),$ the time it takes for a light particle to travel from the stretched horizon to a typical bulk point in the interior of the static patch and back again. One can view the process as the emission of a Hawking particle and subsequent re-absorption  by the horizon.

\item
Beyond $t_c = t_*$ the correlation decreases exponentially with the decay constant of the leading QNM.
 
\end{enumerate}

An obvious goal would be to reproduce these features from correlations of the \dk \ fermions. The simplest possibility would be that the correlation function
\be 
G(t_c) = \frac{1}{N}\sum_i \la \chi_i(0) \chi_i(t_c) \ra
\label{G}
\ee
exhibits behavior similar to figure \ref{bump}.
However this is  not possible. Suppose  $G$ did behave as in figure \ref{corr}. That would mean  an order $1$ fraction of the contributions $\la \chi_i(0) \chi_i(t_c) \ra$ to \eqref{G}   had  similar behavior. In other words a finite       fraction of the $N$  fields $\chi_i$  would have to propagate into the bulk like Hawking modes. Such a large number of fields propagating in the static patch would not be consistent with locality\footnote{Similar issues occur in AdS/CFT where there are $N_{ym}$ fundamental degrees of freedom in an AdS size patch, but only an order $1$ number of bulk fields. In the limit of small $\alpha$ all the bulk string-states fields become degenerate at zero mass but by increasing $\alpha$ the string states separate leading to a sparse spectrum. The analog for \dk \ would be a degenerate spectrum of bulk states for $q\sim 1$ but as $q$ grows the spectrum becomes sparse.}.

Typically we expect the number of propagating fields in de Sitter space to be sparse---of order $1$---certainly not of order $N$. The remaining $\chi$ are important in accounting for the horizon entropy but they must not propagate into the bulk. To put it another way all but a vanishing fraction of the $\chi_i$ should 
be ``confined" to the region very near the horizon and
have correlations which are much shorter range in time, which for very large $N$ would mean that \eqref{G} should have short range behavior.

\begin{figure}[H]
\begin{center}
\includegraphics[scale=.5]{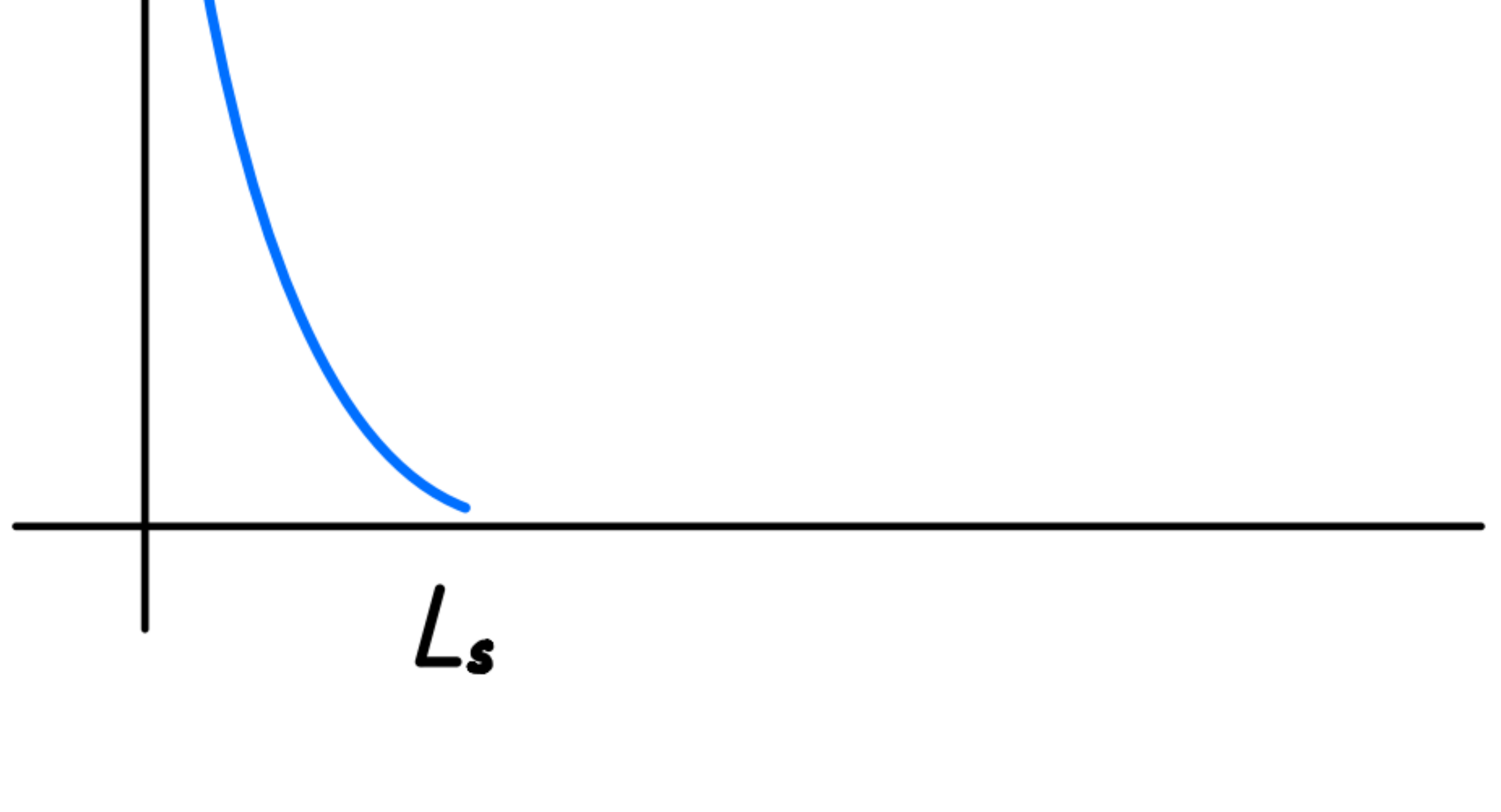}
 \caption{Correlation function for particles confined near the horizon.}
\label{drop}
\end{center}
\end{figure}

Consider the possibility that the number of independent  modes in the bulk is small---of order one---and the rest are trapped near the horizon\footnote{I am grateful to Edward Witten for a crucial discussion on the material in this section.}. The trapped modes would have correlations that persist for a time $\l,$ but since this would be almost all $N$ modes, the ensemble average would be swamped by the trapped modes. Thus
we should expect the ensemble-averaged two-point function to behave like,
\be 
G(t_c) \sim e^{-\CJ t_c}.
\label{chch}
\ee
as illustrated in figure \ref{drop},
\bn
with at most a tiny admixture (that goes to zero in the limit $N\to \infty$) of the behavior in figure \ref{bump}. That is exactly what we will find.

\subsection{DSSYK$_\infty$ Correlator}

\bn

In string units for $\CJ t_s \leq 1$ the correlation function $\la \chi(0) \chi(t_s) \ra$ is known to have the form \cite{Roberts:2018mnp},
\bea
G(t_s) &=& \lf \frac{1}{\cosh^2{\CJ t_s}} \rg^{1/q} \cr \cr
&\to& e^{-\CJ |t_s|/q}
\label{mismtch}
\eea
Equation \eqref{mismtch}  is not a sensible representation of the correlation,   but not surprisingly in cosmic units 
 the it takes the sensible form,
\be 
\la \chi(0) \chi(t_c) \ra \sim e^{-\CJ |t_c|}.
\ee
 This agrees well with the idea (see section \ref{S:FSL}) that  the operators of small size are related to cosmic scales.

 The correlation persists for a short cosmic time $\sim \l$ agreeing with figure \ref{drop},  indicating that most particles are trapped near the horizon and do not propagate deep into the static patch.  The fraction which do propagate into the bulk,  appearing as Hawking radiation, is vanishingly small.  Isolating these propagating \dof \ is an important goal but I will leave it for future work.

\bn

 \sc
 \section{String Scale and the Flat Space Limit} \label{S:FSL}
 
Cosmic scales are  mainly about very low energy quanta of wavelength $\sim \l$  and energy $1/\l$.  Such quanta would be composed of one, or a small number of fermions. Schematically we denote the operators that create them by   $\chi^n$  where $n\sim 1.$ These low energy particles belong to the cosmic sector since their wavelengths are of cosmic scale.

On the other hand the typical energy/momentum of particles in the string range is  $M_s = q\CJ$. The momentum-size relation
 \cite{Susskind:2018tei}\cite{Susskind:2019ddc}  implies that such objects are  composed of $\sim q$ fermions. A simple model might be $q$ weakly bound fermions created by an operator of size $\sim q.$ We will denote the associated  operators schematically by $ \chi^q $.

The ensemble average of correlation function of two such operators has the form \cite{Roberts:2018mnp},
\be  
\la \chi^q \chi^q \ra = \frac{1}{\cosh^2{\CJ t_s}}.
\label{chichi}
\ee
By contrast with \eqref{mismtch} this is a sensible function of string time; all dependence on $q$ and $N$ have canceled out.  This is just as it should be for operators describing string scale excitations.

Operators of size $\sim q$ have recently been the subject of investigation \cite{Berkooz:2018jqr}\cite{Lin:2022rbf}. There is a very rich mathematical theory making use of ``chord diagrams" and quantum Liouville theory. It is not my purpose to discuss this theory here but merely to point out that whatever  it is, it is the theory of string-scale excitations in the SCL.

\subsection{Interactions}
To study the string-scale  in a bit more detail we can consider the interactions between a pair of objects deep, in the interior of the static patch. The objects are  composed of $Q_1$ and $Q_2$ fermions,  with $Q_i \sim q$.

 In general dimensions the potential energy between objects scales  (power-law scaling) with the masses of the objects and the distance between them. In this respect $(2+1)$-dimensions is an exception. The potential energy scales with the  masses but is only logarithmic in distance. The following crude estimate ignores this dependence.

\begin{figure}[H]
\begin{center}
\includegraphics[scale=.4]{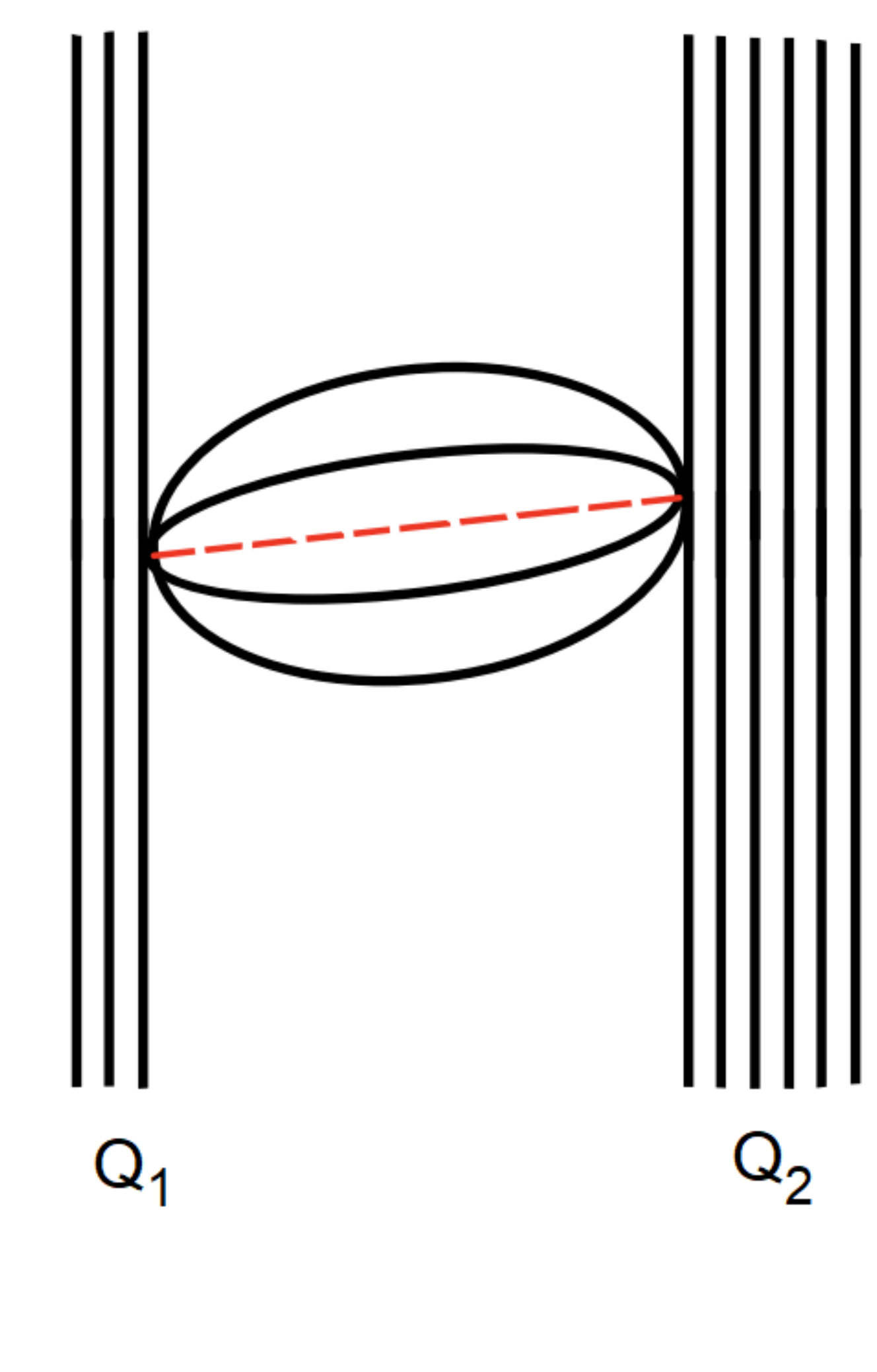}
\caption{Interactions between two clusters of size $\sim q.$}
\label{QQ}
\end{center}
\end{figure}

The dominant graph controlling the potential energy between them $Q_1$ and $Q_2$  is shown in figure \ref{QQ}.
As a perturbative diagram it is IR divergent but section \ref{S:strscale} explained how to regulate the divergence by replacing $\int dt$ with the renormalized expression $$\int \frac{dt}{\cosh^2{q\CJ t}} \sim \frac{1}{\CJ q}.$$
The result is,
\be  
V = Q_1 Q_2 \CJ^2 \frac{N^{q-2}}{(q-2)!}     \frac{q!}{N^{q-1}} \times \frac{1}{\CJ q}
\label{V=QQ}
\ee

We expect that the masses of the two objects are approximately $M_1=Q_1\CJ, \ M_2=Q_2\CJ.$ Equation \eqref{V=QQ}    can be written,
\bea 
V &\sim&  q  \frac{M_1 M_2 }{N\CJ}  \cr \cr
\eq qM_1 M_2 G
\eea
Most of this expression is familiar with the exception of the factor of $q$ which makes the formula not sensible according to  footnote 6.
But the factor $q$ is easily accounted for when we recall that we are working in cosmic units. $V$ is an energy and to convert it to string units we need to divide by $q.$ In string units the formula is sensible,
\be 
V \sim M_1 M_2 G.
\label{Vscale}
\ee

Equation \eqref{Vscale} is the $(2+1)$-dimensional scaling for the Newtonian gravitational potential between the two objects. Note that in the final expression is sensible (in micro or string units)  as all dependence on $q$ and $N$ have canceled out.  (A detailed calculation would also have to produce the logarithmic dependence on the distance between the objects in $(2+1)$-dimensional de Sitter space but that is beyond the scope of this paper.)

\subsection{Why are the Fermions Confined?}

\begin{figure}[H]
\begin{center}
\includegraphics[scale=.4]{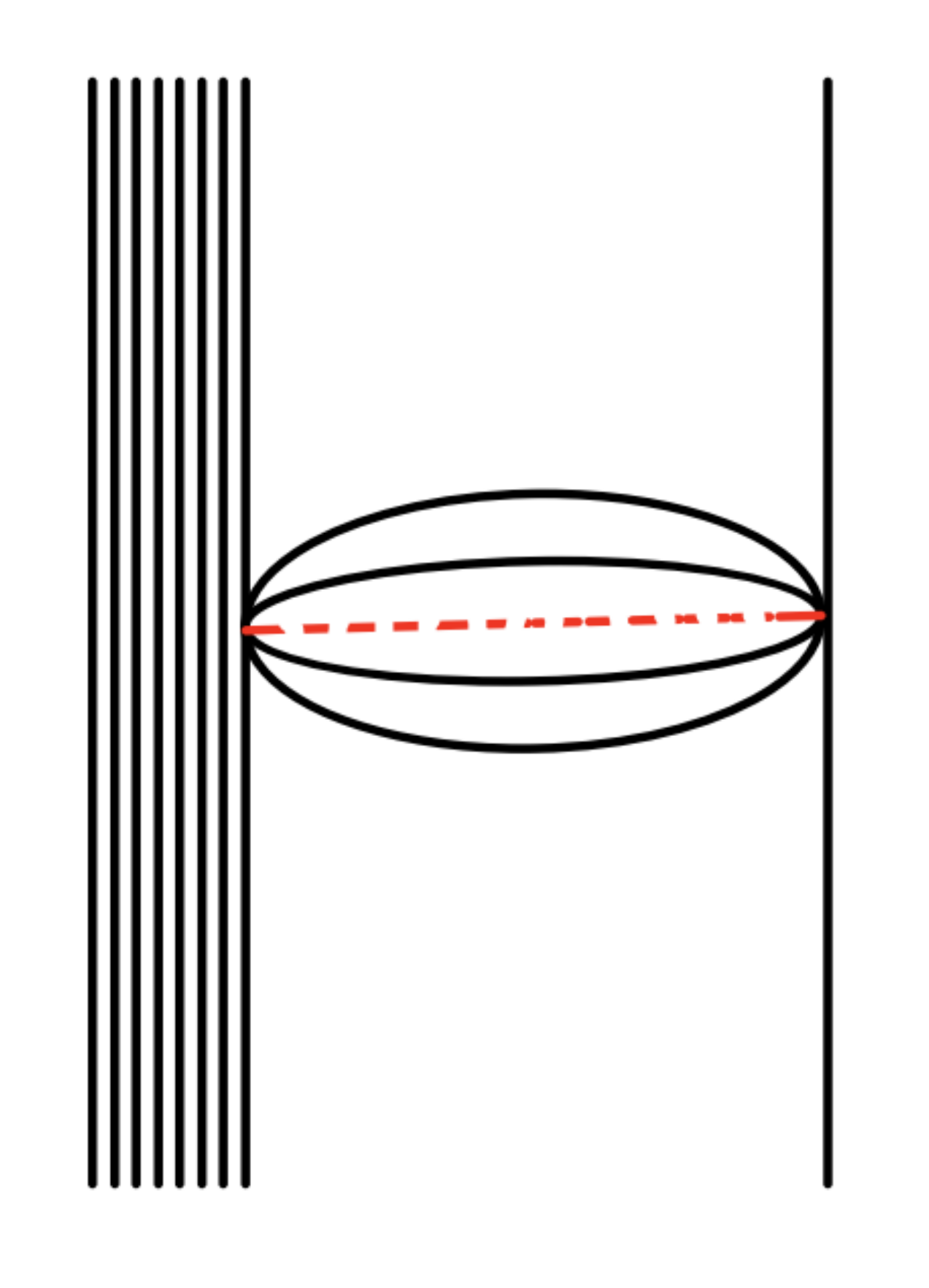}
\caption{Interaction between a single fermion and the remaining horizon 
degrees of freedom.}
\label{confine}
\end{center}
\end{figure}

What is the mechanism that confines the fermions to the immediate vicinity of the horizon, as exemplified by \eqref{chch} and figure \ref{drop}? I suspect that the answer is actually a form of confinement that takes place at the string scale.  The origin of the effect  lies in the interaction of single fermions with the large number of horizon degrees of freedom. That interaction  
(illustrated in figure \ref{confine}) is a special case of the mechanism described in figure \ref{QQ} and equation \eqref{V=QQ}, in which $$Q_1=1, \ Q_2 = N.$$
Notice that in this case $Q_2$ is not of order $q$ but is  much larger, as befits the horizon.
\bn
 Plugging   $Q_1=1, \ Q_2 = N.$  into \eqref{V=QQ} we find that the interaction increases the  energy of the fermion by an amount 
\be 
\Delta E_f = q\CJ,
\label{DE=qJ}
\ee
i.e., the string scale energy. This is much larger than $\CJ,$ the original bare energy of the fermion.
One possible explanation for this effect is that each fermion is connected to the horizon by a string.

 More generally figures like \ref{QQ} and \ref{confine} represent the exchange of string scale systems (they contain $q$ fermions) which  might as well be called strings.

If this is correct then the question is not what confines the fermions, but why isn't any collection of fermions confined? In other words  how do the small number of degrees of freedom that constitute Hawking particles (if there are any\footnote{I don't know any reason why there have to be propagating light fields in three dimensional de Sitter space. In general the only such field which is guaranteed to exist is the graviton, but in $(2+1)$-dimensions there are no gravitons.}) escape? I will return to this question in the near future.

\sc
\section{The Algebra of Observables}

Witten \cite{Witten:2021jzq} and  Chandrasekaran, Longo, Penington and Witten \cite{Chandrasekaran:2022cip}   have argued for an algebraic characterization of of the semiclassical limit of de Sitter space.  The analyses in these papers apply only to cosmic scales and have nothing to say about microscopic or string scales.  

According to  \cite{Chandrasekaran:2022cip} the algebra of cosmic observables in the SCL   is a Type $II_1$ von Neumann algebra, which means that it is based on a maximally mixed density matrix with a flat entanglement spectrum. That condition is met in \dk \ by virtue of the infinite temperature.

From that starting point one builds a Hilbert space by applying finitely many simple operators  (single fermion operators in \dk). The Hilbert space $\CH$ is basically the span of all such states, although there are technical rules for how to complete it by taking appropriate limits. The operator algebra $\CA$ is composed of the bounded operators acting on
 $\CH$.

One  claim of \cite{Chandrasekaran:2022cip} is that the (one-sided) Hamiltonian is not in the algebra. But this raises the question: which Hamiltonian, the cosmic Hamiltonian $i\partial_{t_c}$ or the string Hamiltonian $i\partial_{t_s}$? They are of course proportional to one another for finite $N$ and $q,$ 
$$H_c = q H_s,$$ 
but in the SCL the factor $q$ is infinite.

Let us first consider the algebra of cosmic operators, i.e., operators associated with wavelengths $\sim \l$ and energy $\CJ$. This includes single fermion operators and small number of fermions.  The  energies associated with these operators are of order the tomperature $\CT = \CJ,$ and are characteristic of monomial operators $\chi^n$ with $n\sim 1.$ It is natural to base the von Neumann algebra of cosmic scale operators on the application of finitely many fermion operators as described above and in \cite{Witten:2021jzq}\cite{Chandrasekaran:2022cip}.

But what about phenomena at the string scale? These phenomena involve particles of wavelength $\l/q$ (momenta $\sim q \CJ$) and it is natural to assume that the algebra associated with the string scale is the algebra spanned by monomials $\chi^q$  with $q\sim \sqrt{N}.$ (See for example 
\cite{Berkooz:2018jqr}\cite{Lin:2022rbf}).
Keeping in mind that $q$ diverges in the SCL these operators seem to be distinct from the ones in the cosmic algebra. That's not surprising since the wavelengths of cosmic operators  like $\chi^n$ are outside the flat-space range while the string-scale operators are in that range.
 So it seems that there must be two algebras, $\CA_c$ and $\CA_s$. 

Now  consider a  question raised by the algebraic theory  \cite{Witten:2021jzq}\cite{Chandrasekaran:2022cip}: Are the  Hamiltonians  in their respective algebras; is $H_c$ in $\CA_c$? Is $H_s$ in $\CA_s$?
By definition the double-scaled Hamiltonian, whether it is normalized like $H_c$ or $H_s$  is built from operators of size $q$. 

Intuitively it  seems clear that  $H_c$ cannot be in the algebra $\CA_c,$ since operators of size $q,$ such as $H_c,$  have infinite size in the double-scaled limit, but $\CA_c$ is built of small operators.  

On the other hand there is no obvious obstruction to $H_s $ being in $\CA_s.$ Both  $H_s$ and the operators in $\CA_s$ have size $\sim q.$

To test if an operator belongs in an algebra we follow
 \cite{Witten:2021jzq}\cite{Chandrasekaran:2022cip}  and consider the fluctuation of the operator in the maximally mixed density matrix. If the fluctuation diverges as $N\to \infty$ then the operator cannot be in the algebra.

First consider the cosmic Hamiltonan. We wish to compute $\la H_c^2 \ra.$  This is straightforward and is given by the diagram in figure  \ref{melon}, but with  no integral, and $t_2=t_1$.The result is,
\be 
\la H_c^2 \ra = \CJ^2 N.
\ee
 This is divergent for large $N.$ Therefore $H_c$ is not in $\CA_c$, in agreement with the intuitive argument above as well as with 
 	\cite{Witten:2021jzq}\cite{Chandrasekaran:2022cip}. 
 
 But now consider $\la H_s^2 \ra.$ Recall that $H_s = \frac{1}{q} H_c.$ Thus,
 \bea
 \la H_s^2 \ra &=&   \CJ^2 \frac{N}{q^2 } \cr \cr
 \eq \frac{ \CJ^2}{\lambda}.
 \eea
By definition $\lambda$ is fixed and finite  in \dk \  so $H_s$ passes the test to be in the algebra $\CA_s.$  
To summarize, things are not as simple as just saying that the Hamiltonian is (or is not) in the algebra of observables.  As claimed in  \cite{Witten:2021jzq}\cite{Chandrasekaran:2022cip} 
the cosmic Hamiltonian $H_c$ is not in the algebra $\CA_c,$ but on the other hand\footnote{This observation is due to Henry Lin, private communication.} the string Hamiltonian is in the algebra $H_s.$

The analysis  of \cite{Chandrasekaran:2022cip} is correct as far as it goes, but it is also incomplete. It applies only to the cosmic theory but the cosmic observables are only half the story, the other half being the micro or string-scale observables. The operators describing these observables have size $\sim \sqrt{N}$ and, as far as I can tell without being expert at von Neumann algebras,  do not belong to the algebra $\CA_c.$ To accommodate them it seems possible that the SCL algebra must be expanded to include a ``string sector" $\CA_s$ composed of operators of size $\sim \sqrt{N}.$

\bn

\subsection{Implications?}

I am not sure of the importance of $H_c$ existing or  not existing in the cosmic algebra. As  emphasized in
\cite{Witten:2021jzq}\cite{Chandrasekaran:2022cip} the boost operator exists since the divergences cancel in $(H_{cL} - H_{cR}).$ This should guarantee  that one-sided cosmic evolution (in a static patch) can be defined, using the boost to generate time-translation. But if the individual one-sided Hamiltonians do not exist then $(H_{cL} + H_{cR})$ also does not exist in the algebra. 

Why is this a problem? One concern is  that without $(H_{cL} + H_{cR})$ we may not be able to  evolve (for cosmic times) behind the horizon, since observables behind the horizon are two-sided.

A related  issue:  Does $H$ not being in the algebra imply that $e^{iHt}$ is not defined? That seems wrong in general. For example, the momentum operator does not exist in the Hilbert space of square integrable wave functions but $e^{iap}$ is a perfectly well-defined translation operator for any $a$.

\sc
\section{Summary}
To summarize: The limiting theories---the semiclassical limit of de Sitter space on the gravity side, the $N\to \infty$ on the \dk \ side---appear to be dual to one another, although a sharp proof is still lacking.
 In the limit, on both sides of the duality a separation of cosmic and micro scales takes place. The cosmic scale involves wavelengths of order the de Sitter radius; too large to be treated in the flat-space limit. Micro scales involves shorter wavelengths which are too short to feel the cosmic length scale $\l,$  but not so short that the corresponding energy is enough to back-react significantly on the global geometry. In the SCL the separation becomes so extreme that the two sectors decouple:  one cannot tell from the micro physics that globally  the  cosmological background is de Sitter. To see that the two sectors originate from a common fundamental holographic theory
 one must go slightly beyond SCL and then compare with a more  comprehensive theory that can extrapolate to finite $N.$ 
 
 This paper shows that as the semiclassical and  large $N$ limits are approached the separation of scales takes place in a parallel way in the gravitational theory and its conjectured holographic dual, \dk. Whether or not we regard this as evidence for the duality (I do), it is certainly something we have to take into account, and perhaps take advantage of,  in future analyses of holographic de Sitter dualities.

At the present time the decoupling of scales  applies directly to the recent studies of the limiting micro or string side of \dk \ \cite{Berkooz:2018jqr}\cite{Lin:2022rbf}). Those studies, which take place in the $N\to \infty$ limit, show no evidence of a de Sitter cosmology. But they shouldn't---not without $1/N$ corrections. By the same token the properties of the cosmic limit---tomperature, hyperfast scrambling, etc---show no evidence of any particular micro or string structure.
But when taken together, the string and cosmic phenomena  discussed in this paper imply a common origin. Moreover \dk \ has computable $1/N$ corrections which  couple the two scales and show they
 fit into a common  framework based on \dk.

\section*{Acknowledgements}
I wish to thank Henry Lin,  Geoff Penington,  Adel Rahman, Douglas Stanford,  Edward Witten, Zhenbin Yang, and
 Ying Zhao for critical discussions.


\begin{thebibliography}{99} 



\bibitem{Gibbons:1977mu}
G.~W.~Gibbons and S.~W.~Hawking,
``Cosmological Event Horizons, Thermodynamics, and Particle Creation,''
Phys. Rev. D \textbf{15}, 2738-2751 (1977)
doi:10.1103/PhysRevD.15.2738


\bibitem{Susskind:2021esx}
L.~Susskind,
``Entanglement and Chaos in De Sitter Space Holography: An SYK Example,''
JHAP \textbf{1}, no.1, 1-22 (2021)
doi:10.22128/jhap.2021.455.1005
[arXiv:2109.14104 [hep-th]].

\bibitem{Susskind:2022dfz}
L.~Susskind,
``Scrambling in Double-Scaled SYK and De Sitter Space,''
[arXiv:2205.00315 [hep-th]].

\bibitem{Lin:2022nss}
H.~Lin and L.~Susskind,
``Infinite Temperature's Not So Hot,''
[arXiv:2206.01083 [hep-th]].

\bibitem{Adel}
A. Rahman,``dS JT Gravity and Double-Scaled SYK", To Appear Simultaneously



\bibitem{Witten:2021jzq}
E.~Witten,
``Why Does Quantum Field Theory In Curved Spacetime Make Sense? And What Happens To The Algebra of Observables In The Thermodynamic Limit?,''
[arXiv:2112.11614 [hep-th]].





\bibitem{Chandrasekaran:2022cip}
V.~Chandrasekaran, R.~Longo, G.~Penington and E.~Witten,
``An Algebra of Observables for de Sitter Space,''
[arXiv:2206.10780 [hep-th]].




\bibitem{Cotler:2016fpe}
J.~S.~Cotler, G.~Gur-Ari, M.~Hanada, J.~Polchinski, P.~Saad, S.~H.~Shenker, D.~Stanford, A.~Streicher and M.~Tezuka,
``Black Holes and Random Matrices,''
JHEP \textbf{05}, 118 (2017)
[erratum: JHEP \textbf{09}, 002 (2018)]
doi:10.1007/JHEP05(2017)118
[arXiv:1611.04650 [hep-th]].


\bibitem{Berkooz:2018jqr}
M.~Berkooz, M.~Isachenkov, V.~Narovlansky and G.~Torrents,
``Towards a full solution of the large N double-scaled SYK model,''
JHEP \textbf{03}, 079 (2019)
doi:10.1007/JHEP03(2019)079
[arXiv:1811.02584 [hep-th]].


\bibitem{Lin:2022rbf}
H.~W.~Lin,
[arXiv:2208.07032 [hep-th]].



\bibitem{Maldacena:1997re}
J.~M.~Maldacena,
``The Large N limit of superconformal field theories and supergravity,''
Adv. Theor. Math. Phys. \textbf{2}, 231-252 (1998)
doi:10.1023/A:1026654312961
[arXiv:hep-th/9711200 [hep-th]].



\bibitem{Maldacena:2016hyu}
J.~Maldacena and D.~Stanford,
``Remarks on the Sachdev-Ye-Kitaev model,''
Phys. Rev. D \textbf{94}, no.10, 106002 (2016)
doi:10.1103/PhysRevD.94.106002
[arXiv:1604.07818 [hep-th]].




\bibitem{Roberts:2018mnp}
D.~A.~Roberts, D.~Stanford and A.~Streicher,
``Operator growth in the SYK model,''
JHEP \textbf{06}, 122 (2018)
doi:10.1007/JHEP06(2018)122
[arXiv:1802.02633 [hep-th]].


\bibitem{Susskind:1998vk}
L.~Susskind,
``Holography in the flat space limit,''
AIP Conf. Proc. \textbf{493}, no.1, 98-112 (1999)
doi:10.1063/1.1301570
[arXiv:hep-th/9901079 [hep-th]].

\bibitem{Polchinski:1999ry}
J.~Polchinski,
``S matrices from AdS space-time,''
[arXiv:hep-th/9901076 [hep-th]].

\bibitem{Susskind:2018tei}
L.~Susskind,
``Why do Things Fall?,''
[arXiv:1802.01198 [hep-th]].


\bibitem{Susskind:2019ddc}
L.~Susskind,
``Complexity and Newton's Laws,''
Front. in Phys. \textbf{8}, 262 (2020)
doi:10.3389/fphy.2020.00262
[arXiv:1904.12819 [hep-th]].

















\end{thebibliography}
\end{document}